\documentclass[aps,pre,floatfix,twocolumn,nofootinbib]{revtex4-2}

\usepackage{amssymb,amsmath,amsfonts}
\usepackage[pdftex]{graphicx} 
\usepackage{microtype}    
\usepackage{braket}
\usepackage[usenames]{color}
\usepackage{bm}

\newcommand{\ethr}{e_\text{thr}}
\newcommand{\emarg}{e_\text{marg}}
\newcommand{\emax}{e_\text{max}}
\newcommand{\tsol}{t_\text{sol}}

\begin{document}

\title{Entropic barriers as a reason for hardness in both classical and quantum algorithms}

\author{Matteo Bellitti}
\affiliation{Department of Physics, Boston University, Boston, MA 02215, USA}

\author{Federico Ricci-Tersenghi}
\affiliation{Dipartimento di Fisica,
Sapienza Universit\'a di Roma,
P.le A. Moro 2, I-00185 Roma, Italy}
\affiliation{CNR, Nanotec, Rome unit,
P.le A. Moro 2, I-00185 Roma, Italy}
\affiliation{INFN, Sezione di Roma I,
P.le A. Moro 2, I-00185 Roma, Italy}

\author{Antonello Scardicchio}
\affiliation{The Abdus Salam International Center for Theoretical Physics, Strada Costiera 11, 34151, Trieste, Italy}
\affiliation{INFN Sezione di Trieste, Via Valerio 2, 34127 Trieste, Italy}

\begin{abstract}
    We study both classical and quantum algorithms to solve a hard optimization problem, namely 3--XORSAT on 3--regular random graphs.
    By introducing a new quasi--greedy algorithm that is not allowed to jump over large energy barriers, we show that the problem hardness is mainly due to entropic barriers.
    We study, both analytically and numerically, several optimization algorithms, finding that entropic barriers affect in a similar way classical local algorithms and quantum annealing.
    For the adiabatic algorithm, the difficulty we identify is distinct from that of tunnelling under large barriers, but does, nonetheless, give rise to exponential running (annealing) times.
\end{abstract}

\date{\today}

\maketitle

\section{Introduction}
\label{sec:introduction}

Hard discrete optimization problems are ubiquitous in scientific disciplines and practical applications. The problem of minimizing a complex cost function (or equivalently maximizing a reward function) naturally appears in many different contexts: e.g.\ in physics in the computation of ground state configurations, in statistics in the maximization of the likelihood, in machine learning in the training of artificial neural networks, and so on.

Although real world problems have usually local structures that make their analysis difficult, it is commonly believed that the main source of computational hardness arises from the strong long--range correlations that exist among variables, and this effect can be studied also in more idealized and simple-to-solve models. In other words, in hard optimization problems, starting from an optimal or near-optimal configuration, the change of a single variable (or a small subset of variables) often requires the rearrangement of many more variables in order to remain close to optimality; often the variables to be rearranged are not even close to the modified variable. This property makes the search for the optimal configuration a challenging task even for sophisticated algorithms (see for example the case of backtracking algorithms, like DPLL \cite{arora2009computational}).

An ideal setting for studying this kind of hard optimization problems is provided by constraint satisfaction problems defined on sparse random graphs. Such problems have a twofold benefit: they can be solved analytically using the cavity method, a tool from statistical physics of disordered systems, and they can be efficiently handled on a computer, as the finite mean degree of the graph makes the computational resources required (CPU and memory) grow only linearly with the problem size.

Random constraint satisfaction problems (rCSP) are optimization problems where $N$ discrete variables need to be assigned in order to satisfy $M=\alpha N$ constraints, each one involving a small subset of variables. 
The most famous among rCSP is maybe random $K$-SAT \cite{hartmann2005phase}.
Recently these rCSP have been the subject of intense studies based on statistical physics ideas with the aim of understanding the origin of their computational hardness \cite{ricci2001simplest,mezard2002analytic,barthel2002hiding,mezard2003two,cocco2003rigorous,montanari2004instability,mezard2005clustering,mertens2006threshold,krzakala2007gibbs,krzakala2007landscape,montanari2008clusters,zdeborova2007phase,altarelli2008relationship,krzakala2016statistical}. Indeed, a common feature of all the hard rCSP is the presence of a broad range of the constraints per variable ratio $\alpha$ such that solutions to the problem exists with high probability (in the large $N$ limit), but all known solving algorithms are unable to find any solution in a time growing polynomially with the problem size $N$.
In this hard region it is expected that any solving algorithm requires a time growing exponentially with the problem size, $t \sim \exp(a N)$.

By defining an energy function that counts the number of violated constraints, one can visualize the rCSP as the problem of searching for a zero-energy configuration in a complex energy landscape. The hard phase in rCSP does actually corresponds to an energy landscape with exponentially many (in the system size) local minima that can trap the searching dynamics.
The energy barriers between these minima are usually considered the main source of computational complexity, as any local dynamics is required to jump over these barriers in order to proceed further in the search for the optimal configuration.

Based on the above picture, it is often believed that a quantum evolution --that allows for tunneling events-- may escape local minima more efficiently than a classical stochastic dynamics. This physically reasonable expectation implies that quantum algorithms may be faster in the search for optimal configurations than their classical counterparts, and has fueled interest in quantum algorithms that could benefit from this phenomenon: Quantum Annealing \cite{farhi2000quantum,santoro2006optimization,farhi2012performance,altshuler2010anderson,farhi2014quantum,laumann2015quantum}, its more recent variant, the Quantum Approximate Optimization Algorithm \cite{farhi2014quantum}, and Population Transfer \cite{mossi2017ergodic,mossi2017transverse,smelyanskiy2020nonergodic} are some well-known examples. All these algorithms typically show a complexity growth comparable with the best classical algorithms but, despite their initial promise, it is entirely possible that the limitations they exhibit are insurmountable, and it is unlikely that they could solve NP-hard problems in polynomial time. The limitations might come from the exponentially small tunneling rate out of a local minimum (a phenomenon linked to many-body localization and the existence of an emergent integrable dynamical phase \cite{basko2006metal,altshuler2010anderson,laumann2014many,imbrie2017local}) or might come from other dynamical phenomena. In this paper we identify one such phenomenon.

We consider one of the hardest sparse rCSP, namely random 3-XORSAT, and show that the problem hardness can be interpreted as coming essentially from \textit{entropic barriers}: we introduce a quasi--greedy algorithm, unable to jump over large energy barriers, and notice that it is able to solve the problem as efficiently as the state-of-the-art algorithms, which are designed to be efficient in problems with large energy barriers. This peculiar property suggests that this model is a perfect candidate to understand the effect of entropic barriers. 

We investigate several algorithms, both classical and quantum, in order to better understand the effect of entropic barriers. For all the algorithms analyzed, we find that the time to reach a solution scales exponentially with the system size and quantum dynamics seem to suffer from the presence of entropic barriers as much as the classical algorithmic dynamics. As the effort to build a quantum computer are finally giving up some results \cite{arute2019quantum}, we believe it is important to identify all possible stumbling blocks for quantum architectures.


\section{Model definition and its known solution}
\label{sec:model}

The random 3-XORSAT problem is among the simplest rCSP\cite{ricci2001simplest}: it is made of $N$ binary variables $x_i\in\{0,1\}$  that have to satisfy $M=\alpha N$ parity checks of the kind
\begin{equation}
x_i \oplus x_j \oplus x_k = b_{ijk}\;,
\end{equation}
where the variables entering each constraint are randomly chosen and the parity check bit $b_{ijk}$ is 0 or 1 with equal probability.  In a $K$--XORSAT problem each constraint involves $K$ variables, but for the sake of simplifying the presentation we restrict to the random 3--XORSAT problem, where each constraint involves exactly 3 randomly chosen variables.
Increasing $\alpha$ the typical problem becomes more and more difficult to solve. Solutions exist with high probability in the large $N$ limit until the sat--unsat threshold $\alpha_s$ \cite{dubois20023}. However, the most interesting transition from the point of view of searching algorithms is the clustering (or dynamical) transition that takes place at $\alpha_d$ before $\alpha_s$ \cite{mezard2003two}.
For $\alpha\in[\alpha_d,\alpha_s]$ the space of solutions is shattered in exponentially many cluster of solutions \cite{mezard2003two,cocco2003approximate} and this is what makes the search for solutions much more difficult \cite{ricci2001simplest,mezard2003two,altarelli2008relationship}. This picture has been proven rigorously to a large extent \cite{ibrahimi2012set}.
The hardness of the problem of finding a solution for some classes of local algorithms in the region $\alpha\in[\alpha_d,\alpha_s]$ has been proven \cite{gamarnik2019overlap}, and found to depend on the so-called `overlap gap property', which in practice corresponds to the clustering of solutions taking place after the dynamical transition $\alpha_d$.
In other words, for $\alpha>\alpha_d$ the geometry of solutions is such that the Hamming distance $d$ between any pair of solutions is either very small $d<d_1$ (for pairs of solutions in the same cluster) or very large $d>d_2$ (for pairs of solutions in different clusters) \cite{mezard2003two}. It is exactly the existence of such a range of distances with no solutions that creates an algorithmic bottleneck, while for $\alpha<\alpha_d$ local algorithms can sample the space of solutions \cite{ibrahimi2012set}. So we are going to focus our attention on a 3-XORSAT problem beyond the dynamical threshold $\alpha_d$.\footnote{Strictly speaking, XORSAT is in P, as the solution to a set of $\alpha N$ linear equations in $N$ variables can be always found in time $O(N^3)$ by a simple Gaussian elimination algorithm (more complicated algorithms can do marginally better). 
However, such an algorithm is very special to XORSAT and cannot be generalized to other CSPs being inherently non-local (i.e.\ the operations performed involve in general many variables, also very distant on the interaction network). On the contrary, when \emph{local} algorithms are run on XORSAT they are found to be very slow and affected strongly by the dynamical transition at $\alpha_d$. Even the very same Gaussian elimination algorithm has an algorithmic phase transition at $\alpha_d$ from linear to cubic behavior \cite{braunstein2002complexity} and it does not work as soon as the problem is slightly perturbed (e.g.\ by the addition of a tiny fraction of constraints with the OR operator).}

Given that we are interested in using this model as a benchmark for optimization, we need two more ingredients: (i) We need to define an energy function whose ground state configurations are the solutions to the problem; the simplest choice consists in just counting the number of violated parity checks via the following Hamiltonian
\begin{equation}
\mathcal{H}_0[\bm{s}] = \frac12 \left(M -\sum_{a=1}^M J_a \prod_{i\in\partial a} s_i\right)\;,
\end{equation}
where $s_i=(-1)^{x_i}$ are Ising spins and $J_a=(-1)^{b_a}$ the couplings, being $a$ an index running over all interactions (triplets for 3-XORSAT) and $\partial a$ the set of variables entering the $a$-th interaction.
(ii) At least a solution must always exist, and this can be ensured by enforcing a specific configuration to satisfy all the constraints. For example, by setting all $b=0$, the configuration $x_i=0\;\forall i$ is always a solution. One may think this way of building the model naturally favors the imposed or planted configuration, but this is not the case for the XORSAT problem. As noticed since Ref.~\cite{franz2001ferromagnet}, finding the imposed or planted solution is like finding the crystal in a model of a liquid that upon cooling spontaneously forms a glass: it is well known that crystallization requires an activated dynamical process (nucleation), which is exponentially rare in models with long range interactions, as the random XORSAT. Planted models which are hard to solve are the most natural candidates for optimization benchmarks and the planted XORSAT turns out to be the hardest among these  \cite{barthel2002hiding}. Properties of planted models have been studied in a great detail \cite{krzakala2010following,zdeborova2010generalization} and reviewed in \cite{zdeborova2016statistical}.

The Hamiltonian for the planted model simplifies to the following
\begin{equation}
\label{eq:planted_hamiltonian}
\mathcal{H}[\bm{s}] = \frac12 \left(M -\sum_{a=1}^M \prod_{i\in\partial a} s_i\right)\;,
\end{equation}
which is indeed minimized by the configuration $s_i=1\;\forall i$. This is the energy function we are going to minimize in order to test classical and quantum optimization algorithms.

The last relevant choice regards the interaction hyper-graph, that is the set of $M$ triplets. In the model where the $M$ triplets are chosen randomly the degree of each variable is a Poisson random variable of mean $3\alpha$.
A different choice is the one where the interaction hyper-graph is chosen such that each variable has the same degree $d$: this is called a random regular hyper-graph and can be generated via the configurational model where $N$ variables are given $d$ legs each and $M=Nd/3$ interactions are given 3 legs each, and then variables and interactions legs are coupled in a random way, just avoiding that the same variable enters more than once in the same interaction.
We are going to use the random regular version for the numerical simulations, while the random Poisson version is used for some analytic computations.
The two versions share the same physical behavior.

The statistical properties of the random regular XORSAT problem are well known \cite{franz2001exact,montanari2003nature,montanari2004cooling,krzakala2010following}. Hereafter we are going to focus our studies on the $d=3$ case. In this case the random and the planted models are equivalent in the large $N$ limit for any positive temperature (at $T=0$ the sub-extensive differences between the two models may lead to some discrepancy in the number of solutions discussed in Refs.~\cite{jorg2010first,bapst2013quantum}). We are going to use the planted model in order to be sure that a solution always exists even for finite (and small) values of $N$.

Having fixed the degree of the hyper-graph such that the XORSAT problem is in its hard phase, we can consider now the Gibbs measure corresponding to Hamiltonian $\mathcal{H}$ in Eq.~\ref{eq:planted_hamiltonian} at any temperature $T$. The XORSAT problem corresponds to the problem of finding a zero energy ground state, so it is somehow related to the $T=0$ physics of the model, but the behaviour of the model at $T>0$ is interesting as well.
Indeed it is well known \cite{franz2001ferromagnet} that when a model is beyond the dynamical transition point at $T=0$ (e.g.\ for $\alpha>\alpha_d$) it undergoes a dynamical phase transition at a positive temperature $T_d$ and for $T<T_d$ it has an exponentially large number of metastable states dominating the thermodynamics, $\mathcal{N} \sim \exp[N\Sigma]$, where $\Sigma$ is the so-called complexity.

This is the case for the 3-regular 3-XORSAT model that shows a dynamical phase transition at $T_d=0.255$  and a non-zero complexity of states at $T=0$ that extends from $e\equiv\langle \mathcal{H} \rangle / N = 0$ to $e=e_d=0.0206705$. Actually not all these states are expected to play a relevant role in the relaxation dynamics searching for ground states: from previous studies \cite{montanari2003nature,montanari2004cooling} we expect states above the marginal energy $\emarg=0.018203$ to unlikely trap smart searching algorithms.

Unfortunately a precise connection between relaxation algorithms searching for low energy configurations (e.g.\ $T=0$ Langevin dynamics) and the energy landscape that we can describe in a precise way via the computation of the complexity is still missing (and recent results have clarified that the situation is much more complicated than previously expected \cite{folena2020rethinking,mannelli2020marvels,mannelli2020thresholds}). So we cannot make an analytical claim about the threshold energy which is hard to go below by searching algorithms, but this threshold energy is certainly positive and close to $\emarg$. Reaching a solution, that is an $e=0$ configuration, is a very hard problem and requires in general times scaling exponentially with the system size $N$.

\section{The optimization algorithms}
\label{sec:algo}

\begin{figure}
\includegraphics[width=0.7\columnwidth]{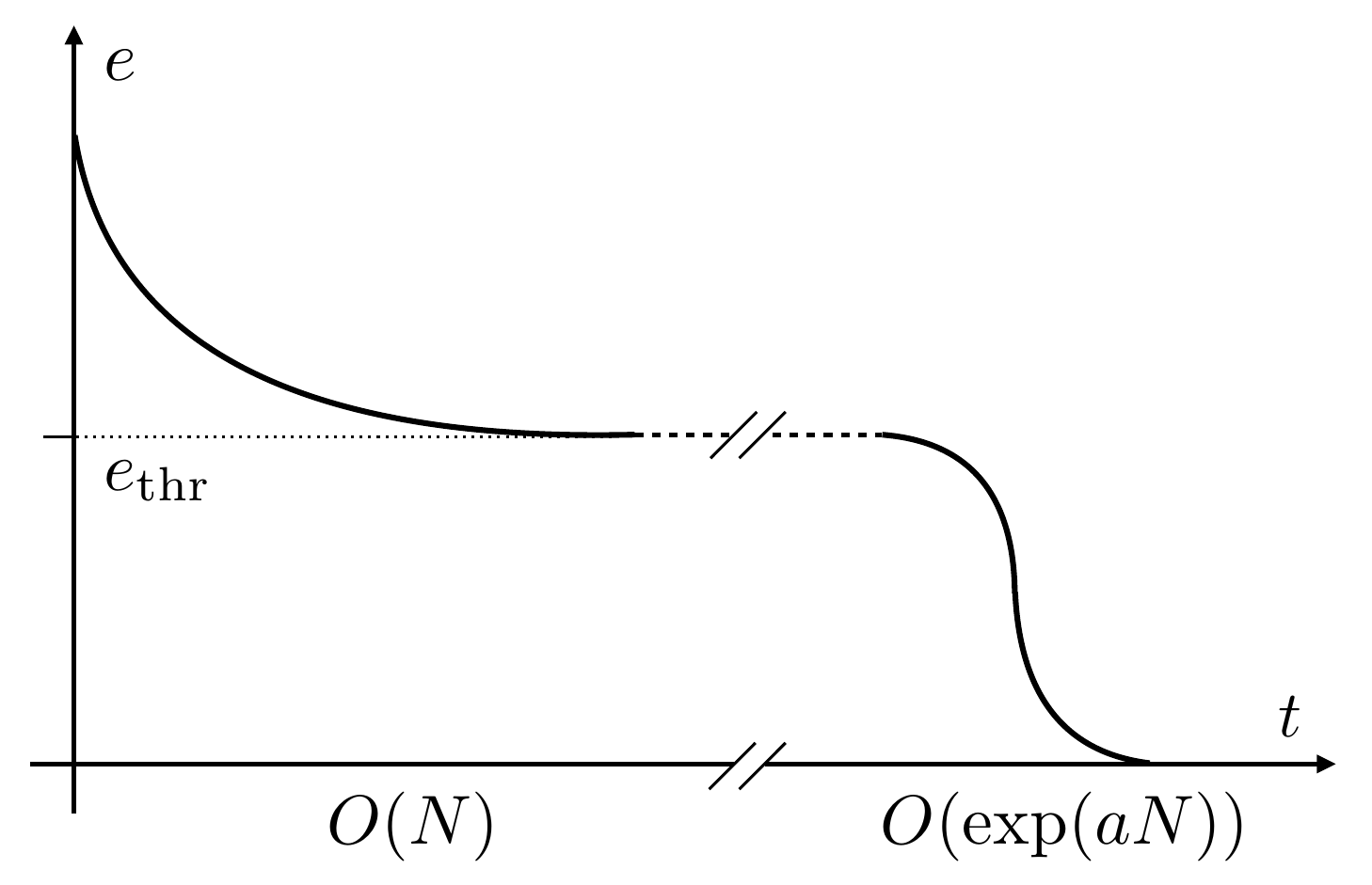}
\caption{Schematic picture for the energy relaxation in hard optimization problems. On short times the energy relaxes to a threshold value, while the ground state (solution) is reached only for times growing exponentially with the system size. The first regime can be described by ordinary differential equations taking the large $N$ limit at $t/N$ fixed, while the second requires the estimation of rare and large fluctuations.}
\label{fig:2regimes}
\end{figure}

In the study of the optimization algorithms, that is the out of equilibrium processes that try to minimize the energy, the order in which the large size limit ($N\to\infty$) and large times limit ($t\to\infty$) are taken is extremely important. We expect an algorithm-dependent threshold energy, $\ethr>0$, to exist such that configurations with $e>\ethr$ can be reached in an ``easy'' way (e.g.\ in a time scaling linearly with $N$), while to reach a solution (i.e.\ a configuration with $e=0$) a time growing exponentially in $N$ is required in general for hard problems.
(see the schematic picture in Fig.~\ref{fig:2regimes}).

We will see that for some algorithms we are able to provide an approximate description of the dynamics in the regime where the $N\to\infty$ limit is taken before the $t\to\infty$ limit, thus estimating $\ethr$ (that in the best cases is close to the marginal energy $\emarg$.)
However the interesting question about the scaling of times to reach a solution requires a different analytic approach where fluctuations are taken into account.
Most of our results are in the regime where times are made large while keeping $N$ finite are based on numerical experiments.

The presentation of our results about optimization algorithms is somehow split in two parts. In Sections \ref{sec:SA} and \ref{sec:QG1} we discuss the regime of linear times where large sizes can be studied and analytical solutions in the large $N$ limit can be obtained (thus estimating the threshold energy for various algorithms). In Sections \ref{sec:walksat} and \ref{sec:QG2} we study the regime where times are made exponentially large in the system size $N$, and estimate the exponential growth rate of the time to reach a solution.

\subsection{Simulated Annealing: a warming up with the most widely used optimization algorithm}
\label{sec:SA}

\begin{figure}
\includegraphics[width=\columnwidth]{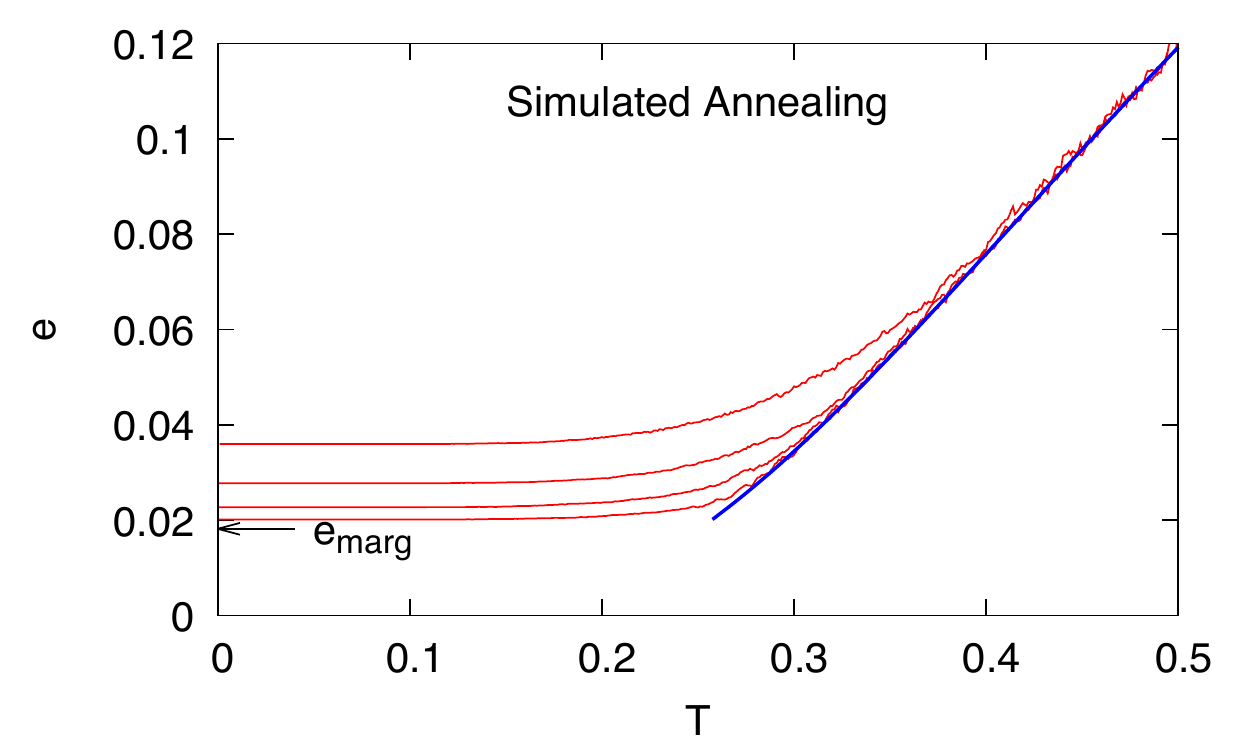}
\caption{Simulated Annealing is not able solve the 3-regular 3-XORSAT problem, since it converges in the long time limit to a positive energy (close to the marginal energy $\emarg$). The four curves have annealing rates  $\Delta T=10^{-3},10^{-4},10^{-5},10^{-6}$ (from top to bottom). The blue curve represents the energy above the dynamical transition.}
\label{fig:SA}
\end{figure}

Simulated Annealing (SA) is maybe the most widely used optimization algorithm. It consists in implementing a Monte Carlo Markov Chain sampling from the Gibbs-Boltzmann distribution $P_\text{GB}({\bm s}) \propto \exp(-\mathcal{H}({\bm s}) / T)$ with a temperature $T$ slowly decreasing towards zero. In Fig.~\ref{fig:SA} we report the results of Simulated Annealing run on samples of size $N=10^5$ with a cooling schedule where the temperature is decreased by $\Delta T$ after each Monte Carlo Sweep (MCS): the four curves corresponds to $\Delta T=10^{-3},10^{-4},10^{-5},10^{-6}$ (from top to bottom).

From the Simulated Annealing results it is clear that an efficient, but still linear-time, algorithm is not able to solve the 3-regular 3-XORSAT problem, and seems to converge to configurations with a threshold energy close to $\emarg$. To achieve the zero energy configuration we need to use an algorithm that can go below the threshold energy, overcoming energetic and/or entropic barriers.

\subsection{A broad class of stochastic search algorithms}
\label{sec:QG1}

Given a particular spin configuration, let us classify its variables according to the number of unsatisfied interactions they belong to: we say a variable is of type $k$ if it belongs to $k$ unsatisfied interactions. In the present model $k\in[0,3]$ since $d=3$ for all variables.
We 	call $f_k(t)$ the fraction of variables of type $k$ at time $t$, that satisfy $0\le f_k(t) \le 1$ and $\sum_{k=0}^3 f_k(t)=1$ at any time.

The searching algorithm we propose is extremely simple and works as follows: at each time step it chooses one variable of type $k$ with probability $p_k(t) \propto w_k f_k(t)$ and flips it. The time is then incremented by $1/N$ in order to have a well defined continuous process in the large $N$ limit.

Starting from a random configuration the behavior of the algorithm is determined only by the vector of weights $\bm{w}=(w_0,w_1,w_2,w_3)$. The stopping condition depends on the weights: if all weights are non-null the algorithm never stops; while if $w_0=0$ the algorithm cannot flip variables participating only in satisfied interactions and thus any solution is a stopping configuration. We fix $w_0=0$ hereafter so as to make any solution a stopping configuration for the algorithm. We study several choices for the vector of weights (the weights need not be normalized, but the probabilities $p_k \propto w_k f_k$ are).

\subsubsection{Analytic description}

Before presenting the actual performance of this algorithm, we would like to stress that the evolution of the algorithm can be described analytically under some assumptions, which are similar to those already used in the literature to approximately describe the relaxation dynamics in model defined on a Bethe lattice \cite{semerjian2003relaxation,cocco2003approximate,semerjian2004approximation}.

At each time step the fractions $\{f_k(t)\}$ change depending on the variable chosen. For example, if a variable of type $k=3$ is chosen and flipped, then that variable changes its type from 3 to 0 ($3\to 0$), i.e.\ the fraction $f_3$ decreases by $1/N$ and the fraction $f_0$ increases by $1/N$; at the same time its 6 neighboring variables decrease by one the number of their types, and one needs to compute the number of changes $n_{3\to 2}$, $n_{2\to 1}$ and $n_{1\to 0}$ in order to properly update the fractions $\{f_k\}$ (for example $\Delta f_2 = (n_{3\to 2} - n_{2\to 1})/N$). The three numbers $n_{3\to 2}$, $n_{2\to 1}$ and $n_{1\to 0}$ are random variables distributed according to the multinomial distribution ${\sf Mult}(\{p_{3\to 2}, p_{2\to 1},p_{1\to 0}\}, 6)$. Under the approximation that no correlation exists between the types of neighboring variables once the common interaction is removed (cavity approximation) we can compute the parameters of the multinomial distribution
\begin{equation}
    p_{3\to 2} \propto 3f_3\,, \quad p_{2\to 1} \propto 2f_2\,, \quad p_{1\to 0} \propto f_1\,,
\end{equation}
where the proportionality constant is fixed by $p_{3\to 2}+p_{2\to 1}+p_{1\to 0}=1$.
One more example: if the variable chosen to be flipped is of type $2$, then, apart from the change $2\to 1$, we have that $n_{3\to 2}$, $n_{2\to 1}$ and $n_{1\to 0}$ are distributed according to ${\sf Mult}(\{p_{3\to 2}, p_{2\to 1},p_{1\to 0}\}, 4)$, while $n_{0\to 1}$, $n_{1\to 2}$ and $n_{2\to 3}$ are distributed according to ${\sf Mult}(\{p_{0\to 1}, p_{1\to 2},p_{2\to 3}\}, 2)$, where
\begin{equation}
p_{0\to 1} \propto 3f_0\,, \quad p_{1\to 2} \propto 2f_1\,, \quad p_{2\to 3} \propto f_2\,,
\end{equation}
with again the normalization condition $p_{0\to 1}+p_{1\to 2}+p_{2\to 3}=1$.
So, the variations of the fractions due to a single spin flip are random variables given by
\begin{multline*}
    \Delta f_k(t) = \frac1N\Big[
    n_{k-1\to k}^{(2(3-j))} - n_{k\to k+1}^{(2(3-j))} +\\
    n_{k+1\to k}^{(2j)} - n_{k \to k-1}^{(2j)} +
    \delta_{j,3-k} - \delta_{j,k} \Big] \quad \text{w/prob.}\;\; p_j(t)\;,
\end{multline*}
where the superscript $(d)$ in $n^{(d)}$ refers to the total number of events in the multinomial distribution, and we fix $n_{-1\to 0}=n_{0\to -1}=n_{3\to 4}=n_{4\to 3}=0$. The generalization of the above equation to other random graphs or interaction types is straightforward.
Rescaling the time such that a spin flip happens every $\Delta t=1/N$, and taking the average over a small, but finite, time interval, corresponding to $O(N)$ single spin flips, the stochastic equation above can be converted in the $N\to\infty$ limit to an ordinary first order differential equation in terms of mean values $\mathbb{E}[n_{k\to \ell}^{(d)}]=d\,p_{k\to \ell}$. For simplicity, we keep using the same notation, but now $\{f_i(t)\}$ are not single trajectories, but mean values over the trajectories:
\begin{multline}
f_k^\prime(t) = \sum_{j=0}^3 p_j(t) \Big[ 
        2(3-j) (p_{k-1\to k} - p_{k\to k+1}) +\\
        2j (p_{k+1\to k} - p_{k \to k-1}) + \delta_{j,3-k} - \delta_{j,k} \Big]\;.
        \label{eq:main}
\end{multline}

\begin{figure}[t]
    \includegraphics[width=\linewidth]{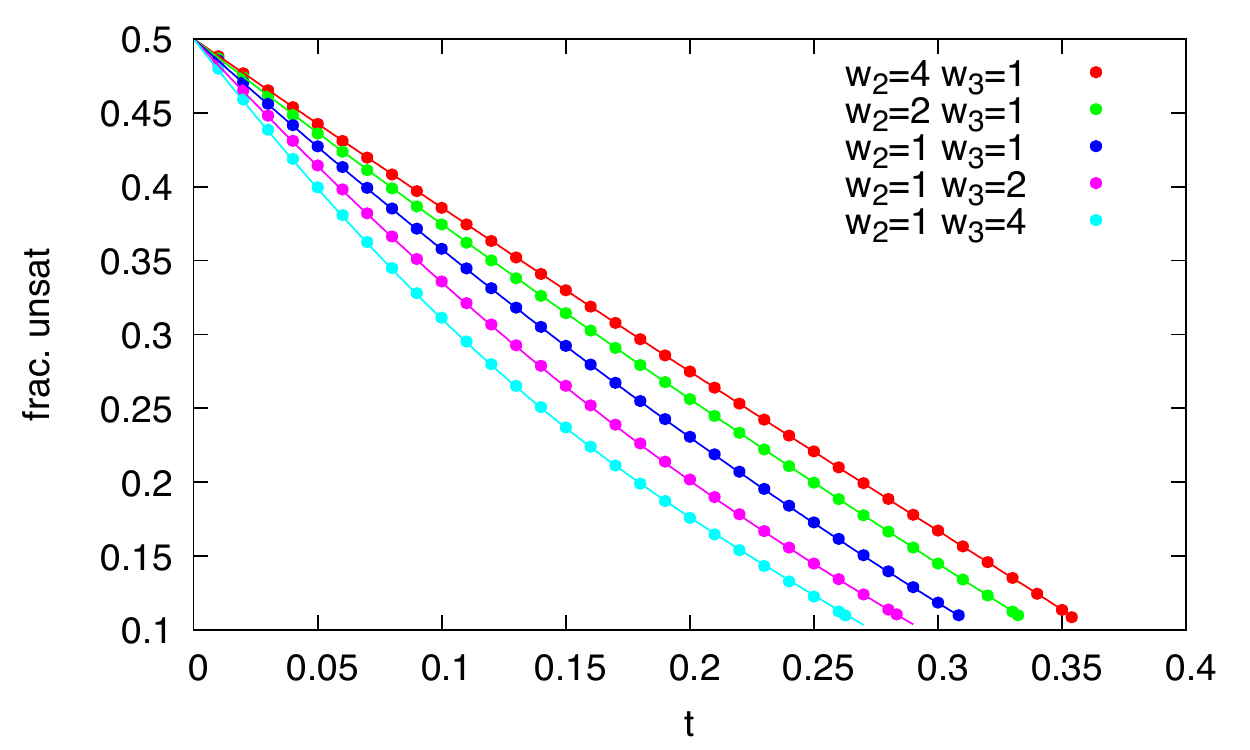}
    \caption{Evolution of the energy for several greedy versions ($w_1=0$) of the algorithm. Points are numerical data and lines are the analytic description.}
    \label{fig:greedyEner}
\end{figure}

An easy check is that $\sum_{k=0}^3 f_k^\prime(t)=0$, so the total probability is conserved. The solution to Eq.~(\ref{eq:main}) can be easily achieved by any integration algorithm for ordinary differential equations. This analytic solution will be compared in the following with the actual evolution of the algorithm.

\subsubsection{Actual performances}

Let us start from a greedy version of the algorithm, that is an algorithm that can never increase the energy (it is the equivalent of gradient descent, but here the space of configurations is discrete, so gradients are not well defined). This greedy version of the algorithm requires $w_1=0$ because flipping a variable of type 1 would increase the energy (i.e.\ the number of unsatisfied interactions); only weights $w_2$ and $w_3$ can be non null in the greedy case. For this greedy version the number of stopping configurations is very large (their entropy is computed in Appendix \ref{app:countingBlocked}): any configuration where each variable participate to 0 or 1 unsatisfied interactions is a stopping configuration. These configurations are local minima of the energy function and we call them \emph{blocked}.

\begin{figure}
    \includegraphics[width=\linewidth]{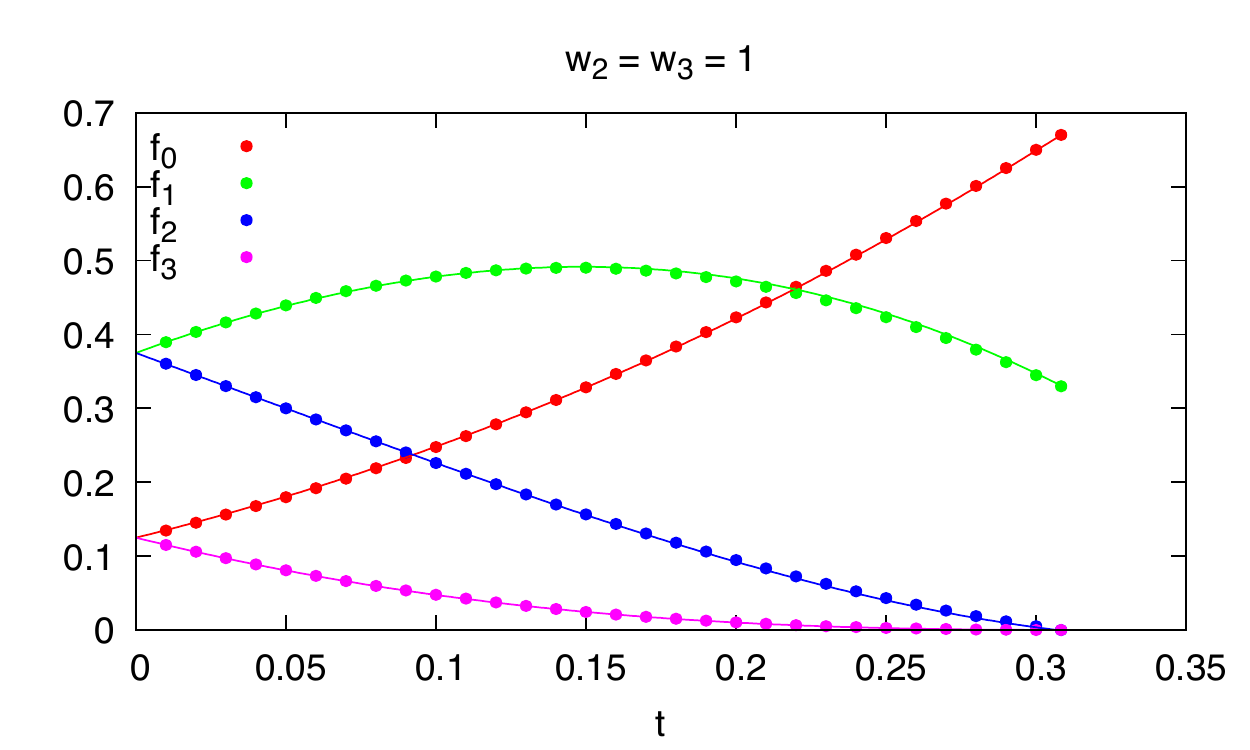}
    \caption{The actual evolution of fraction of variables in a greedy version ($w_1=0$) of the algorithm (data points) are well described by the analytic solution (full curves). The algorithm stops when $f_2=f_3=0$.}
    \label{fig:comp_0_1_1}
\end{figure}

\begin{figure}
    \includegraphics[width=\linewidth]{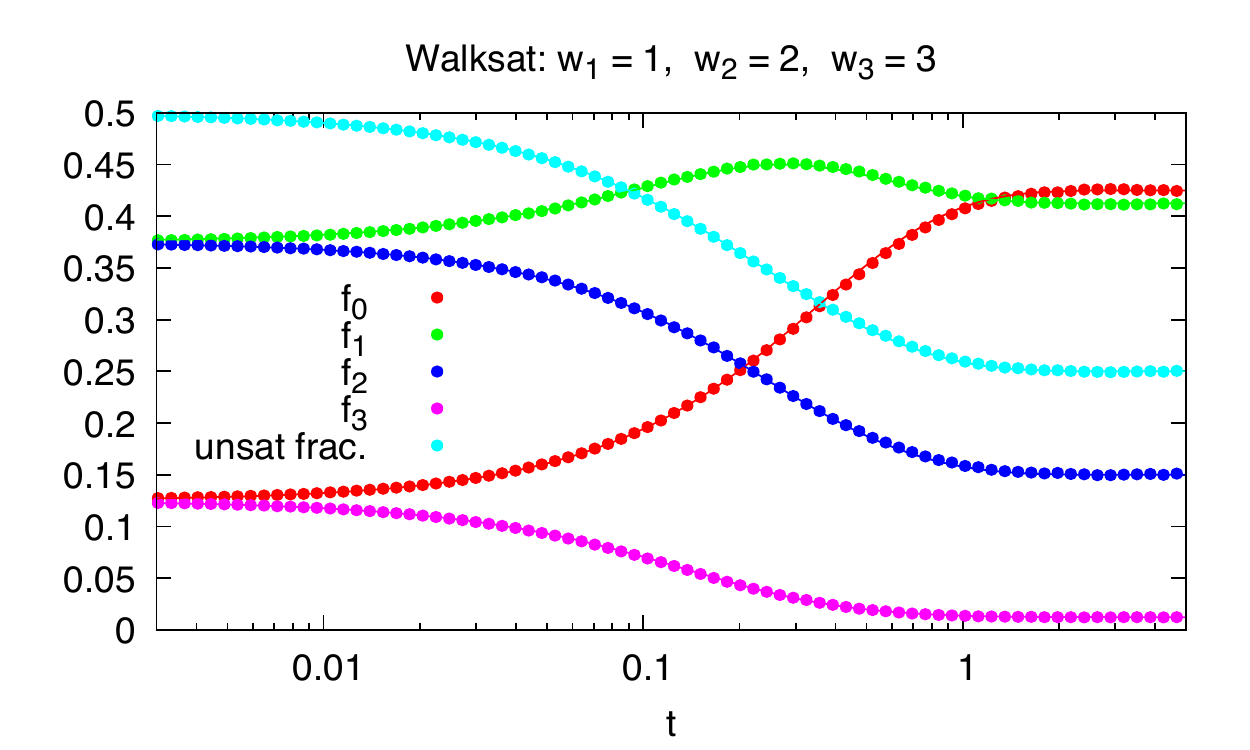}
    \caption{Evolution of the energy and fraction of variables during the WalkSAT algorithm. The analytic description of the relaxation to the threshold energy is almost perfect.}
    \label{fig:walksat}
\end{figure}

\begin{figure*}[tb]
    \includegraphics[width=0.75\linewidth]{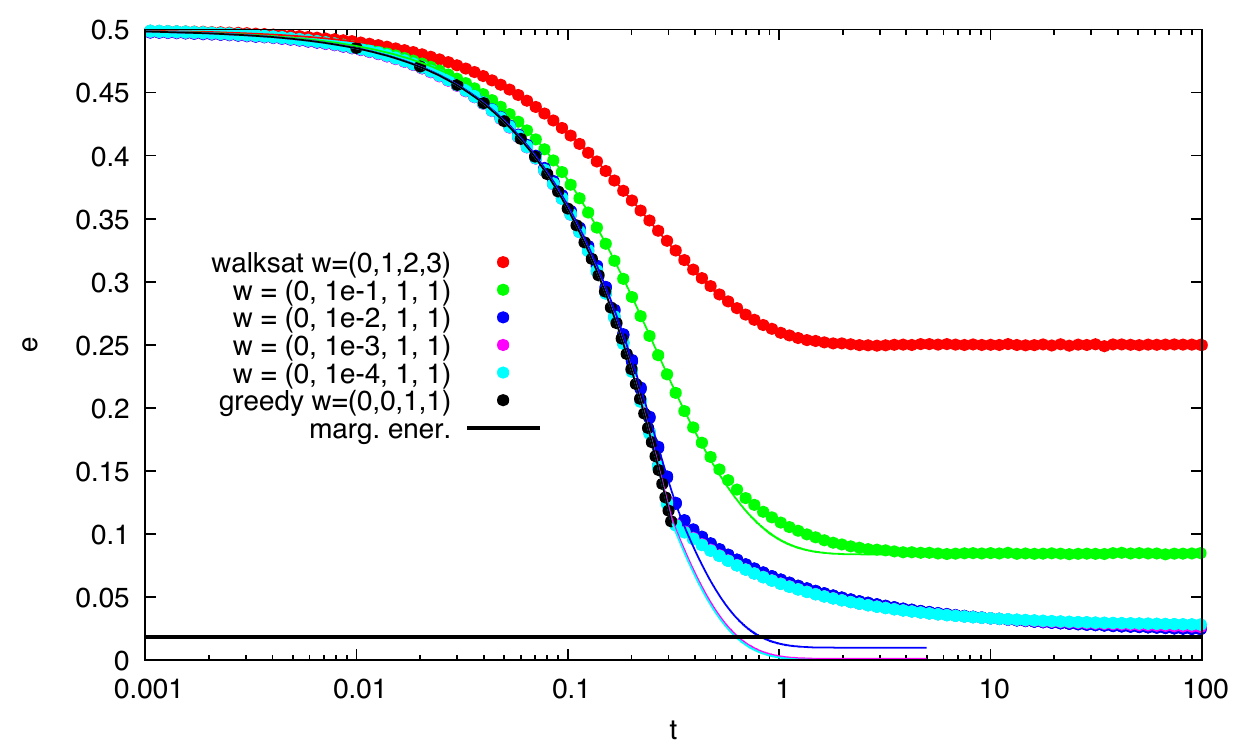}
    \caption{Evolution of the energy for WalkSAT and several (quasi-)greedy versions of the algorithm. The vector of weights $(w_0,w_1,w_2,w_3)$ appears in the legend. The analytic description under the assumption of lack of correlations is reported with a full curve and fails to describe the algorithms at the lowest energies, when correlations arise.}
    \label{fig:quasiGreedy}
\end{figure*}

In Fig.~\ref{fig:greedyEner} we show the energy or fraction of unsatisfied constraints as a function of the running time of the greedy algorithm ($w_1=0$) for different choices of the ratio $w_3/w_2$ (the $w$'s are not normalized so only their ratio matters). Points are the numerical data and the lines are the solution to the ODE in Eq.~(\ref{eq:main}). We notice the agreement between analytics and numerics is almost perfect, because at those energy values correlations are very weak. The same almost perfect agreement can be seen also at the level of fraction of variables $f_k$ participating in $k$ unsatisfied constraints (see Fig.~\ref{fig:comp_0_1_1} for the case $w_2=w_3=1$).

Let us consider now non-greedy versions of this algorithm, that is cases with $w_1>0$ where the energy can sometimes increase during the evolution, although we expect on average the energy to relax to the threshold value.

The first interesting case is the one that corresponds to WalkSAT \cite{selman1993local}. A  detailed description of this algorithm is given in Sec.~\ref{sec:walksat}; for the present purposes it is enough to say that in WalkSAT variables are flipped accordingly to the number of unsatisfied constraints they belong to, that is $w_k\propto k$. The numerical results and the comparison with the analytical description is provided in Fig.~\ref{fig:walksat}: the agreement is excellent. Although this version of the algorithm would run forever, after a finite time a stationary regime is reached and nothing interesting happen any more in the large $N$ limit.
Once the threshold energy has been reached, then the search for the solution takes place via rare fluctuations as discussed in the Sec.~\ref{sec:walksat}.

However, having achieved a good analytic description of the WalkSAT algorithm is not enough, because this algorithm works at very high energies: its threshold energy is $1/4$ for 3-regular 3-XORSAT, more than twice the energy that the greedy version achieves and more than 10 times the marginal energy we expect to be the relevant threshold energy for smart searching algorithms.

So we are interested in studying more efficient versions of this algorithm and we explore the case where $w_1$ is much smaller than $w_2$ and $w_3$ (in practice we fix $w_2=w_3=1$ and explore the range $w_1\ll 1$).
Notice that if $w_1$ is made very small, the algorithm in practice is allowed to make an energy-increasing spin flip only when no other move is allowed, that is when the configuration is blocked. We have simulated also a version of the algorithm where this is made explicit and the results are equivalent.

In Fig.~\ref{fig:quasiGreedy} we report the evolution of the energy for WalkSAT, the greedy version and several quasi--greedy versions of the algorithm. We observe that the threshold energy is $1/4$ for WalkSAT, close to 0.1 for the greedy version, but approaches very closely the marginal energy $\emarg$ for $w_1\ll 1$. So the quasi--greedy versions of this algorithm seems very effective in reaching the same energy values Simulated Annealing can achieve.

The analytic description of the algorithm changes a lot depending on the energy value. For high enough energies, correlations are very weak and the analytic solution matches perfectly the numerical data: this is true for WalkSAT and the greedy version as already shown in Figs.~\ref{fig:greedyEner}, \ref{fig:comp_0_1_1} and \ref{fig:walksat}. For $w_1=0.1$ we start seeing some deviations at times $t\sim 1$, that luckily enough disappear for larger times.

The most interesting cases are those with $w_1 \le 10^{-2}$. First of all we observe a very weak dependence on $w_1$ (the three values used $w_1=10^{-2},10^{-3},10^{-4}$ give practically the same results), so we are working in the limit where the algorithm does an energy-increasing spin flip only when no other move is available. So this algorithm is not able to jump over barriers of height larger than 1 (or few units at most), that correspond to $\Delta e=O(1/N)$.

Indeed this quasi--greedy algorithm follows closely the greedy version until the time, $t^*\sim 0.3$, when the latter reaches a blocked configuration that stops it. For $t>t^*$ the quasi--greedy algorithm keeps decreasing the energy at a much slower pace, eventually approaching an energy very close to the marginal one (that we expect to be a lower bound for the threshold energy in linear time algorithms).

In the regime $t>t^*$ the analytical approximation based on the lack of correlations fails dramatically in many important aspects: it keeps predicting a very fast energy decrease and it estimates a too small asymptotic energy. The reason for such a failure is clear: low energy configurations are very peculiar and have strong correlations among variables, even among variables which are far apart. In this energy range, our approximation fails and the analytic solution is meaningless.

It is worth stressing that we are still working on timescales growing linearly with $N$ (the first regime depicted in Fig.~\ref{fig:2regimes}). So the failure of the analytic approximation is not due to the fluctuations that become important in the second regime of exponentially large timescales. Here we are still working in a regime such that every spin variable has been flipped a \emph{finite} number of times. Nonetheless the evolution of the quasi-greedy algorithm bring the system in configurations correlated up to some distance, such that hypothesis leading to Eq.~\eqref{eq:main} are no longer valid.

From a preliminary study we have understood that in this energy regime the algorithm proceeds by performing collective rearrangements of variables forming a local tree-like structure, and the size of these collective rearrangements seems to grow while approaching the threshold energy (this reminds us a lot what happens in similar problems approaching the dynamical transition \cite{montanari2006dynamics,semerjian2008freezing}). The analytical description of the quasi--greedy algorithm in this low energy regime is deferred to a future work.

We move now to the problem of estimating the large deviation rate to reach a solution via a rare fluctuation from the threshold energy. Given the differences in the threshold energies clearly visible in Fig.~\ref{fig:quasiGreedy} we expect quite different rates for WalkSAT and the several quasi--greedy versions of the algorithm.

\subsection{WalkSAT: the large deviations rate to reach a solution from a simple argument}
\label{sec:walksat}

WalkSAT is a popular randomized (or stochastic) algorithm for solving constraint satisfaction problems \cite{selman1993local}. In its simplest version works as follows: starts from a random configuration; at each time step, if all constraints are satisfied returns the solution, otherwise picks uniformly at random an unsatisfied constraint and flips a randomly chosen variable participating that constraint. The flip certainly satisfies the chosen constraint (interaction) but may unsatisfy many other constraints: so the algorithm may increase the energy during the evolution.

An approximated analytic description of this algorithm exists \cite{semerjian2003relaxation,barthel2003solving,hartmann2005phase} from which we learn that for the hard problem we are studying the algorithm relaxes to a positive energy in a linear time and then reaches a solution by a rare fluctuation taking an exponential time.
However it is not clear which kind of barrier the algorithm crosses, given that the energy increase is in principle without bounds.

Actually the analytic description of the WalkSAT algorithm can be made even simpler than in previous approaches \cite{semerjian2003relaxation,barthel2003solving,hartmann2005phase} by noticing that (i) the algorithm stays most of the time on configurations of very high energy where correlations are very weak and (ii) the fluctuation leading to a solution is so rapid that correlations do not arise. 

These insights suggest that keeping track of the energy alone should capture the essential features of the algorithm. In practice, we think of the energy as a stochastic process, and write down a stochastic differential equation that describes its dynamics during the computation. The details of the calculation are presented in the following section.

\subsubsection{Analytic description}
In this section we focus on random $p$-XORSAT, with $\alpha$ constraints per spin: every spin participates in a Poisson distributed number of interactions (with average and variance $p \alpha$), while every interaction involves exactly $p$ spins.
Random $p$-XORSAT is known to be less correlated than the regular version (see the comment after Eq.~\ref{eq:planted_hamiltonian}), so we expect it to be more amenable to a simple description.

Consider a uniformly random spin configuration $\bm{s}$. Every spin is connected to a Poisson distributed number of broken constraints with average $p \alpha_u$, where $\alpha_u = \mathcal{H}[\bm{s}]/N$. The same spin is also connected to a Poisson number of satisfied constraints with average $p (\alpha - \alpha_u)$.
The fundamental idea of our approach is to assume that WalkSAT is incapable of building any correlations that violate this property, which is strictly true only on fully random configurations. This is reasonable, as most of the time is spent in high energy states.

Given an initial random configuration $\bm{s}$, we run WalkSAT for $T$ steps and denote the number of broken constraints at this time $\mathcal{H}(T)$.
When we take one more step, $\mathcal{H}(T)$ changes by
\begin{equation}
\Delta \mathcal{H}_T^{T+1} = -1 -u(T) + s(T)\;,
\end{equation}
where $u(T)$ is the number of \emph{excess} broken constraints connected to the spin (i.e.\ excluding the one that was selected by
WalkSAT) and $s(T)$ is the number of satisfied constraints connected
to it. As explained earlier, we assume that at all times they are distributed as if the configuration was random.

After a number $\Delta T$ of steps, larger than one but
small compared to the number of spins $N$, the total energy change is
\begin{equation}
    \label{eq:De}
    \Delta \mathcal{H}_T^{T+\Delta T} = - \Delta T - \sum_{k=0}^{\Delta T -1}
    u(T) + \sum_{k=0}^{\Delta T -1} s(T)\;,
\end{equation}
and we approximate the two sums using the central limit
theorem: 
\begin{align}
\sum_{k=0}^{\Delta T -1} u(T) &\approx p \alpha_u \Delta T + R_1 \sqrt{p
\alpha_u \Delta T}\;, \\
\sum_{k=0}^{\Delta T -1} s(T) &\approx p (\alpha - \alpha_u) \Delta T + R_2 \sqrt{p (\alpha - \alpha_u) \Delta T}\;,\nonumber
\end{align}
where $R_1$ and $R_2$ are two independent standard Gaussian random variables. In these expressions we assume $\Delta T$ is short enough that the energy density $\alpha_u$ does not appreciably change over the interval $ \Delta T$.
The sum of the two terms involving $R_1$ and $R_2$ is again a Gaussian variable, so that the total energy change after $ \Delta T$ steps is approximately
\begin{equation}
\Delta \mathcal{H}_T^{T+\Delta T} = - \Delta T + p (\alpha - 2 \alpha_u) \Delta T + \sqrt{p \alpha} R \sqrt{\Delta T}\;.
\end{equation}
We are interested in the large $N$ limit, so it is convenient to change variable from $T$ to $t = T/N$. Dividing the previous equation by $N$ we obtain a stochastic differential equation describing an Ornstein-Uhlenbeck process:
\begin{equation}
\label{eq:sde_walksat}
 d \alpha_u(t) = 2 p \left( \frac{p \alpha - 1}{2p} -  \alpha_u
 \right) dt + \sqrt{\frac{p \alpha}{N}} dW\;.
\end{equation}
To estimate the scaling with $N$ of the time necessary to reach a solution, it is best to study the Fokker--Planck equation associated to Eq.~\eqref{eq:sde_walksat}:
\begin{equation}
    \label{eq:fokker_planck} 
    \frac{\partial P}{\partial t} = - \frac{\partial}{\partial \alpha_u} 
    \Big( 2p \left( \frac{p \alpha - 1}{2p} - \alpha_u \right) P \Big) +
    \frac{1}{2} \frac{p \alpha}{N} \frac{\partial^2 P}{\partial
    \alpha_u^2}\;.
\end{equation}
Regardless of the boundary condition $P(\alpha_u,t=0)$, after an initial transient, the energy is distributed according to the stationary solution of Eq.~\eqref{eq:fokker_planck} (the normalization constant $c_N$ is irrelevant here)
\begin{equation}
    P_\text{st}(\alpha_u) = c_N \exp \left(
    - 2 N \frac{ (\alpha_u - \alpha_0)^2}{\alpha}\right)\;,
\end{equation}
which is centered around the finite value (if $p \alpha > 1$)
\begin{equation}
    \alpha_0 = \frac{p\alpha-1}{2p}\;.
\end{equation}
This indicates that the time necessary to reach the solution grows more rapidly than $O(N)$, but also gives us a concrete way to estimate it:
the typical time over which a transition from energy $\alpha_0$ to zero happens is the reciprocal of the rate
\begin{align}
\label{eq:rate_poisson} 
    \tsol & \sim \frac{1}{\text{Prob}(\alpha_u = 0,t \to \infty \ | \ \alpha_u = \alpha_0, t = 0)} \nonumber \\
    &= \exp \left( N \frac{(p \alpha
    -1)^2}{2 \alpha p^2} \right)\;.
\end{align}
As expected, the solution time scales exponentially with the problem size. This analytical solution does show, in its simplicity, the mechanism by which a solution is found and why we call it \emph{entropic barriers}. At equilibrium, reached after $O(N)$ steps, the system fluctuates around a very high energy threshold $\alpha_0$. This is well above any local minimum where the system could be stuck. With exponentially small probability however a rapid fluctuation of $O(N)$ spin flips can bring the system to the real solution, but this happens with exponentially small probability. So, only an exponentially small fraction of all the possible sequences of $N$ spin flips will pick the right direction to the solution, hence the problem is of an entropic nature, not an energetic one. We will discuss this in more details after comparing with numerical results.

\begin{figure}
    \includegraphics[width=\linewidth]{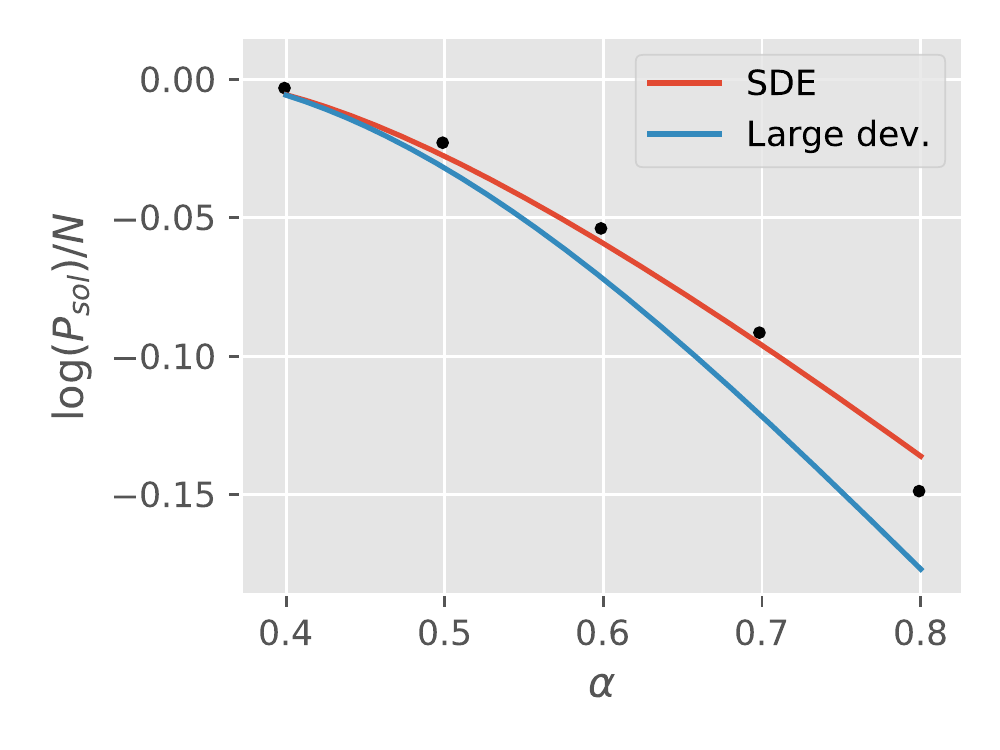}
    \caption{Comparison to the numerics for 3-XORSAT. The vertical axis is the logarithm of the probability of solution in linear time divided by $N$. The black dots are the numerical data reported in
    \cite{semerjian2003relaxation} and the blue line is the analytic approximation derived in that reference, based on large deviation theory. The red line is our approximation (see Eq.\eqref{eq:rate_poisson}).}
    \label{fig:compare_sm}
\end{figure}

\subsubsection{Comparison with numerics}

\begin{figure}
    \centering
    \includegraphics[width=\linewidth]{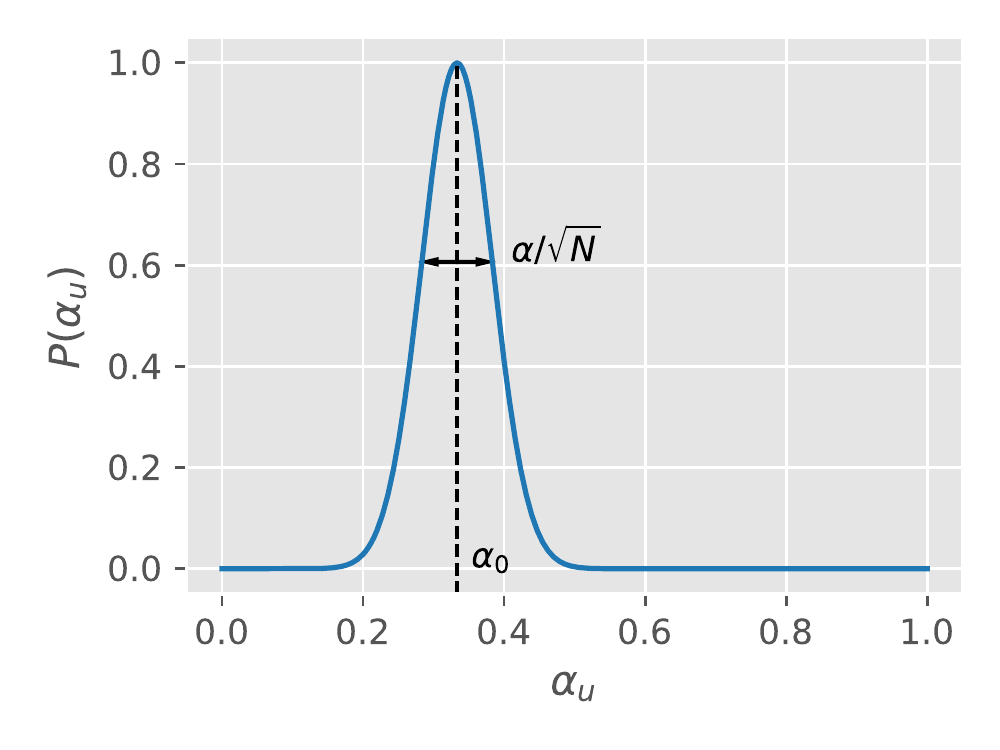}
    \caption{Equilibrium distribution of the density of UNSAT clauses $\alpha_u$ for the Fokker-Planck equation (\ref{eq:fokker_planck}). A solution is found when a rare fluctuation reaches the value $\alpha_u=0$, which occurrs with probability exponentially small in $N$.}
    \label{fig:PDE_equilibrium}
\end{figure}

A comparison with the actual rate measured in numerical experiments in Ref.~\cite{semerjian2003relaxation} is provided in Fig.~\ref{fig:compare_sm}. The quality of the rate reported in Eq.~\eqref{eq:rate_poisson} and obtained under the assumption of lack of any correlation is surprising and supports the idea that WalkSAT makes a random search without building any relevant correlation in the problem.

It is interesting to notice that the same simple argument can be made also in the 3-regular 3-XORSAT case and provides the following scaling for the time to a solution
\begin{equation}
\label{eq:rate_regular} 
    \tsol \sim \exp(N/6)\;.
\end{equation}
The rate $\mu=1/6\simeq 0.167$ is again close to the numerics reported in Ref.~\cite{guidetti2011complexity}, where $\mu \approx 0.124$ was measured.

Notice that the physical interpretation of this calculation is that the equilibrium distribution is centered around a \emph{relatively large} non-zero value of the cost function, as seen in Fig.~\ref{fig:PDE_equilibrium}. The solution is found by one rare fluctuation which brings the system out of equilibrium by a ``lucky'' series of $O(N)$ flips which points in the exact direction to the ground state. During this series, no significant correlation is created between the values of the various random variables so the central limit theorem we used is valid. We expect this to be the behavior of a random algorithm which works at high temperature, tackling a problem with entropic barriers.

In the analysis of more complex algorithms, the ones that work at lower temperatures, this assumption is most probably violated. Correlations between the random variables need to be taken into account when computing the rate of the rare fluctuation that will lead to finding of the ground state. If one wants to make progress in the analysis of such algorithms one needs to go beyond what done in this paper.

\subsection{Times to reach a solution via the quasi--greedy algorithm}
\label{sec:QG2}

Given that the quasi--greedy version of the algorithm has a much lower threshold energy than the WalkSAT algorithm we expect rare fluctuations leading to a solution to happen with a better exponential rate.
Unfortunately in the quasi--greedy case the computation can not be done analytically because, as seen above, the approximation based on the lack of correlations provides very poor results and so it is not useful at all. We thus resort to a numerical computation.

We have observed above that in the limit $w_1\ll 1$ the dependence on $w_1$ is extremely weak in the first regime of linear times. The same is true also for the second regime of exponential times. The exponential rate $\mu$ determining the mean time to reach a solution $\tsol \sim \exp(\mu N)$ does not show any visible dependence on $w_1$.
So, we have fixed $w_1=0.05$ to carry on our numerical experiments with the quasi--greedy version of the algorithm.

\begin{figure}
    \includegraphics[width=\linewidth]{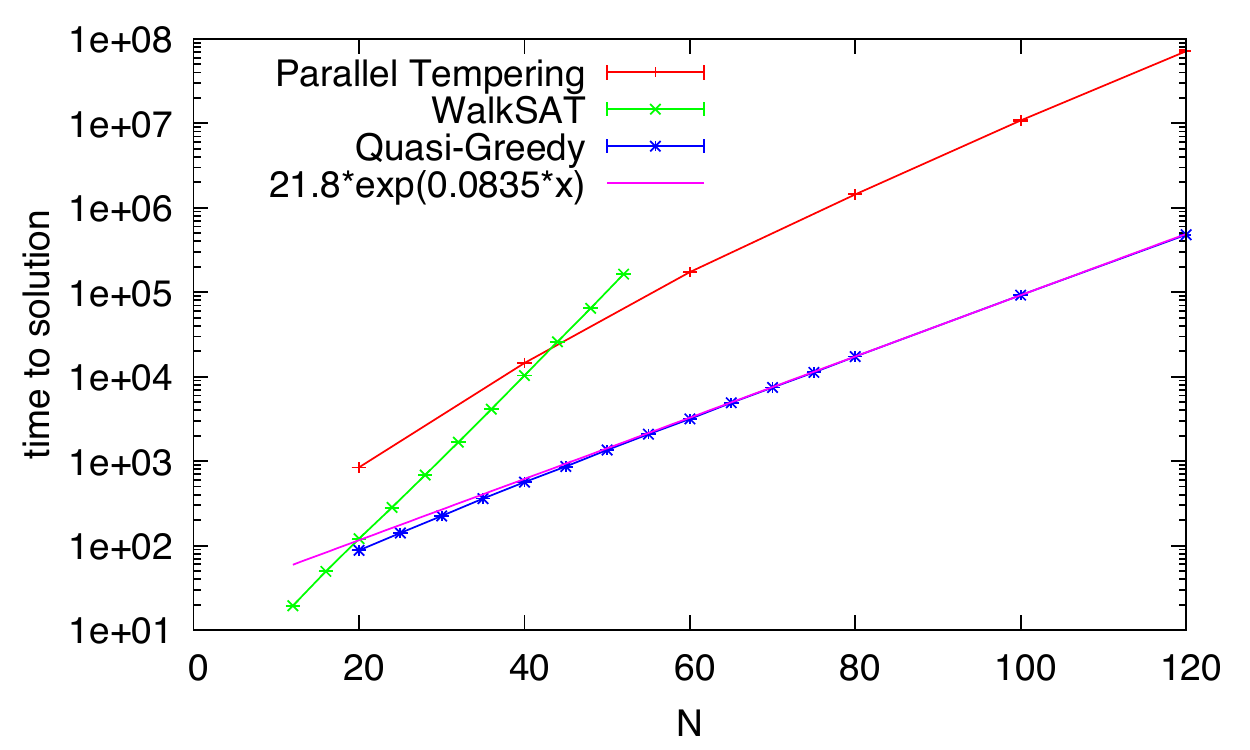}
    \caption{Mean times to reach a solution measured as spin flips per variable. The exponential growth of the time to reach a solution by the quasi--greedy version of the algorithm has a rate $\mu_\text{QG}\approx 0.0835$, much smaller than the one of WalkSAT and comparable to the one of Parallel Tempering (which is however much slower because of the prefactor).}
    \label{fig:tempiMedi}
\end{figure}

In Fig.~\ref{fig:tempiMedi} we report with blue points the average time to reach a solution running the quasi--greedy algorithm. The time we report is the mean number of sweeps, where a sweep corresponds to $N$ spin flips. The average is taken over $10^6$ different runs and the error is smaller than the symbol size. For large $N$ the data is very well fitted by an exponential growth with rate $\mu_\text{QG} = 0.0835(5)$.

In the same figure we report also the mean time to reach a solution by the WalkSAT algorithm. It is clear that the rate is much larger, although for these sizes we are still observing the preasymptotic behavior and for larger sizes the rate will converge to $\mu_\text{WS} \approx 0.124$ \cite{guidetti2011complexity}.

A much more meaningful comparison is the one with the Parallel Tempering (PT) algorithm. PT is considered the state-of-the-art for thermalization (and optimization) in the field of disordered systems, as it can deal efficiently with very rough energy landscapes. For spin glass models, which are similar to the problem we are studying here, it has been recently shown that PT performs comparably to Population Annealing (PA) \cite{wang2015population} another standard algorithm to optimize complex energy functions. We prefer to study the performance of PT in finding a solution, rather than PA, because in PT it is straightforward to compute the time to solution once the temperature scheduling is fixed. While in PA one needs to increase the size of the population during the run and if the solution is not found a new run with larger sizes should be done; moreover comparing population sizes and running times requires some extra care.

We have run PT with an optimal temperatures scheduling derived in Appendix \ref{app:PTscheduling}.
The results in terms of Monte Carlo Sweeps per replica are shown in Fig.~\ref{fig:tempiMedi}. The comparison with the quasi--greedy algorithm is very favorable for the latter: the growth rate for PT is slightly larger than $\mu_\text{QG}$ (but still decreasing and could eventually converge to the same value) and the prefactor for PT is larger by 2 orders of magnitude than the one for the quasi--greedy algorithm.

\subsection{Quantum annealing}
\begin{figure*}[bt]
    \centering
    \includegraphics[width=\linewidth]{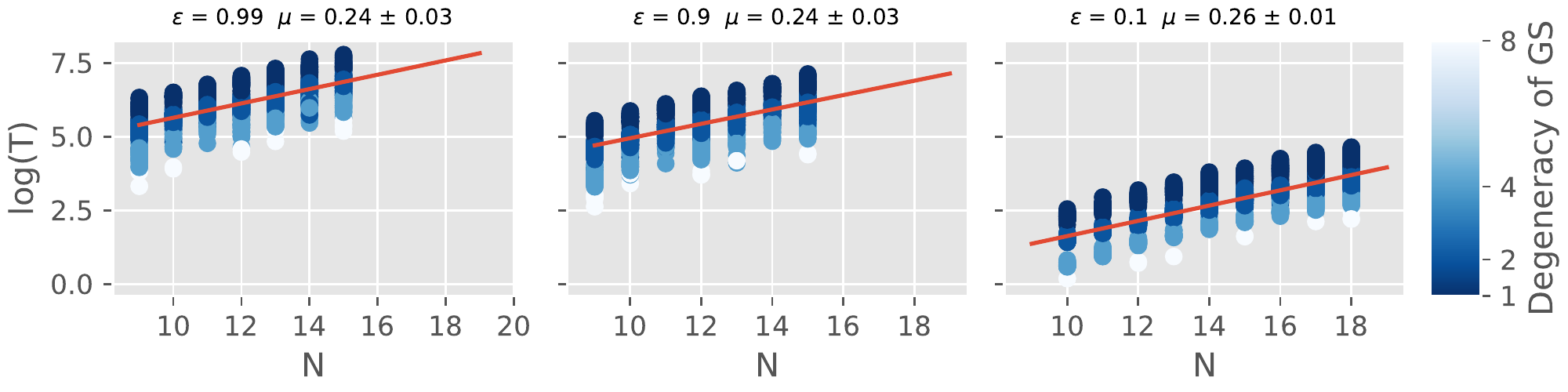}
    \caption{Annealing time $T$ as a function of instance size $N$ and solution probability $\varepsilon$. Every dot corresponds to an instance, with lighter color indicating a more degenerate ground state. Notice how the rate $\mu$ is insensitive to the target probability, as expected from the restart argument given in the main text. The red line is a linear fit performed on the median value of the annealing time, using the standard deviation of $\log(T)$ as error bar.}
    \label{fig:QA_results}
\end{figure*}

Originally proposed twenty years ago \cite{kadowaki1998quantum,farhi2000quantum}, Quantum Annealing is a quantum algorithm designed to solve \emph{classical} optimization problems, exploiting the adiabatic theorem of quantum mechanics \cite{ambainis2004elementary,messiah1962quantum}.
First, we encode the given classical problem in a \emph{problem Hamiltonian} $H_P$, so that solutions correspond to its ground states. The problem Hamiltonian is typically in the form of a cost function with all terms commuting among themselves (in this sense it is a classical problem). A convenient and customary choice is to have the problem Hamiltonian be diagonal on the $\sigma^z$ basis.
Then, we choose a \emph{fluctuation Hamiltonian} $H_F$, an arbitrary operator that should provide ``quantum fluctuations'' and must have a known and simple ground state. A popular choice is to use a uniform field in the $\sigma^x$ direction to provide fluctuations:
\begin{equation}
    \label{eq:unif_trans_field}
    H_F = \sum_{i} \sigma^x_i\;,
\end{equation}
where the sum runs over all spin variables in the system. 
The Quantum Annealing algorithm consists then in time--evolving the ground state of
\begin{equation}
    H(t) \equiv \frac{t}{T} H_P + \left( 1 - \frac{t}{T} \right) H_F 
\end{equation}
from $t=0$ to $t=T$. For a long enough annealing time $T$, the adiabatic theorem guarantees that a system initially prepared in the ground state of $H(0) = H_F$, known by construction, will evolve into a state belonging to the ground state manifold of $H(T) = H_P$. Measuring this state on the $\sigma^z$ basis we obtain a solution to the original problem.
Notice that choosing the fluctuation Hamiltonian as in Eq.~(\ref{eq:unif_trans_field}) guarantees the initial ground state has finite overlap with every state in the computational basis; the algorithm will not miss a solution only because the corresponding state had no initial amplitude.
The adiabatic theorem also provides a lower bound on $T$: for the algorithm to succeed, the annealing time should be longer than
\begin{equation}
\label{eq:adiabatic_theorem}
    T \gg \frac{ \max_t \braket{\psi_1(t)| \partial_s H(s) | \psi_0(t)}}{
    \min_t \Delta^2(t)} \qquad s \equiv \frac{t}{T}\;,
\end{equation}
where $\ket{\psi_0(t)}$ is the instantaneous ground state at time $t$, $\ket{\psi_1(t)}$ is the first excited state, and $\Delta(t)$ is the energy gap between them.
%
%
We have already encountered a classical Hamiltonian encoding XORSAT in Eq.~\eqref{eq:planted_hamiltonian}. 
The quantum mechanical version is simply
\begin{equation}
H_P = \frac{1}{2} \left(N -\sum_{a=1}^N \prod_{i\in\partial a} \sigma^z_i \right)\;.
\end{equation}
In our numerics, we use this as the problem Hamiltonian and a uniform transverse field as the fluctuation one. 
Once we fix the annealing time $T$ and initial state $\ket{\psi(0)}$, QA will end up in some final state $\ket{\psi(T)}$. Let the probability
of measuring an energy equal to the ground state energy, which in our model is $E_\text{GS}=0$, be $\varepsilon$ in this state. 
As $T$ increases, $\varepsilon$ will approach unity. The complexity of the algorithm is not expected to change with $\varepsilon$, as long as $\varepsilon=O(1)$. In fact, since $\varepsilon$ is the probability to find the ground state after a time $T$, one can enhance this probability by repetition of the algorithm. A success probability of $\varepsilon$ after two repetitions becomes $1-(1-\varepsilon)^2=2\varepsilon-\varepsilon^2$, and even a small $\varepsilon$, after $n=O(1/\varepsilon)$ repetitions can be made close to 1. This repetition, or restart technique, is commonly used in algorithms for CSP \cite{hartmann2005phase}, like WalkSAT. 
One commonly adopted definition for the time complexity of QA is to invert the relation between $\varepsilon$ and annealing time: one fixes $\varepsilon$ and $N$, and asks what is the corresponding annealing time $T(N,\varepsilon)$. For large enough $N$ it is reasonable to expect an exponential scaling of the form
\begin{equation}
    T(N,\varepsilon) \sim A(N,\varepsilon) \exp \left( \mu N \right)\;,
\end{equation}
where the prefactor $A(N,\varepsilon)$ is allowed to have a polynomial dependence on $N$, while the rate $\mu$ does not depend on $\varepsilon$. 
In practice, $T(N,\varepsilon)$ is rarely estimated from real--time unitary evolution, since the required computational power quickly becomes unmanageable as the number of spins grows above $N=20$. 
There are two common strategies adopted to sidestep this problem: estimate the energy gap between the ground and first excited state via Exact Diagonalization or Quantum Monte Carlo \cite{jorg2010first,farhi2012performance,bapst2013quantum} and invoke the adiabatic theorem to impose a bound on $T(N,\varepsilon)$, or perform the evolution in imaginary time assuming that the scaling of the imaginary analog of $T(N,\varepsilon)$ will be the same as the original quantity \cite{santoro2006optimization}. Both methods find exponential scaling of the time for hard classical problems, a fact which has been connected with the order of the thermodynamic transition \cite{jorg2010first} (however, see \cite{laumann2012quantum}).
%
%
While both methods provide reasonable bounds on the scaling of the annealing time, it is unclear how the entropy of excited states affects those estimates. Since entropic effects are the primary focus of this work, we decided to perform real--time evolution to minimize confounding factors, even if this means limiting the simulations to moderate sizes.
We integrated the time--dependent Schrödinger equation via an explicit high--order adaptive Runge--Kutta method \cite{dormand1980family} 
implemented in the QuSpin library \cite{weinberg2017quspin}.
%

The results are presented in Fig.~\ref{fig:QA_results}:  it is clear that instances with a unique solution (dark blue dots) are harder than instances with multiple solutions (lighter blue dots), but not exponentially so. For any choice of the solution probability $\varepsilon$, the growth rate $\mu$ is compatible with 
\begin{equation}
    \mu = 0.25 \pm 0.07
\end{equation}
This value is higher than the one predicted in  \cite{farhi2012performance}, that in our notation would read $\mu \simeq 0.167$. The simulations on that reference are run only on instances with a unique solution, and use a Quantum Monte Carlo method to estimate the gap.

\begin{figure}
    \includegraphics[width=\columnwidth]{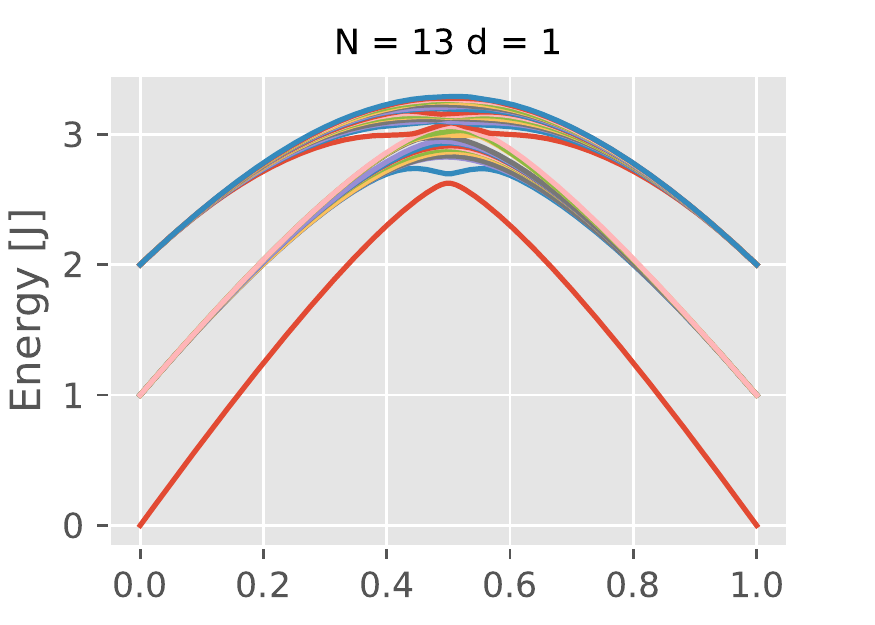}
    \caption{Lowest 30 levels in the spectrum as a function of transverse field strength $t/T$ for an instance with nondegenerate ground state. This corresponds to a dark blue point in Figure \ref{fig:QA_results}.}
    \label{fig:nondegenerategs}
\end{figure}

\begin{figure}
    \includegraphics[width=\columnwidth]{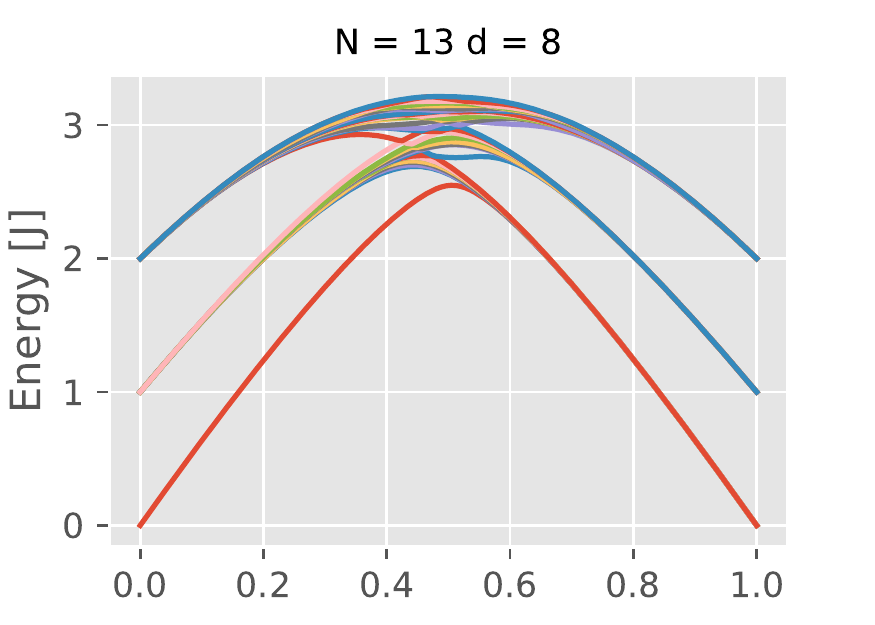}
    \caption{Lowest 30 levels in the spectrum as a function of transverse field strength $t/T$ for an instance with eightfold degenerate ground state. This corresponds to a white point in Figure \ref{fig:QA_results}.}
    \label{fig:degenerategs}
\end{figure}

\section{Conclusions}
\label{sec:conclusions}

In Figs.~\ref{fig:nondegenerategs} and \ref{fig:degenerategs} we compare the spectrum of a typical instance with unique solution to the spectrum of an instance with eight--fold degenerate ground state, as a function of the transverse field. When the ground state is unique (Fig.~\ref{fig:nondegenerategs}), one of the states in the first excited manifold peels off and has an avoided crossing with the ground state. The performance of the Quantum Adiabatic Algorithm is limited by the size of this gap. When the ground state is degenerate (Fig.~\ref{fig:degenerategs}) the picture is qualitatively different: as the transverse field is turned off, a few excited states go through a cascade of avoided crossings, eventually ending in the ground state manifold. In this case, knowledge of the minimum gap between the ground state and the first excited state is not enough to understand the scaling of the annealing time: we should know how many states end up in the ground state, and what avoided crossings those have. Typically, these states will start high in the spectrum, where there will be  many crossings, and so where entropy is important. These instances are not naturally described by the picture of trapped states which tunnel to the solution (like a single level crossing) and we cannot avoid noticing the similarity with the phenomenon of entropic barriers, in the sense that only a ``lucky" sequence of flips from one state to the other can lead to ground state manifold. And this, too, occurs with exponentially small probability.

Last, we must notice that the threshold energy for this problem is $e_\text{marg} \simeq 0.02$, so a more quantitative understanding of the difference between excited states below the threshold energy and above it in the context of Quantum Annealing requires studying instances with $N \simeq 100$ or larger. 
For the small sizes we can currently study via the exact integration of the Schrödinger equation the ground state degeneracy seems the best indicator for identifying hard instances.

For rCSP problems, some algorithms encounter entropic barriers rather than energetic barriers. This in particular happens when the algorithm works at energies higher than the threshold energies for the given problem and they find the ground state by a rare fluctuation which picks the right sequence of $O(N)$ spin-flips among the exponentially many. We have shown this explicitly with a family of almost-greedy algorithms which includes WalkSAT. These algorithms evolve in $t=O(N)$ to their equilibrium state (this part of the evolution can be followed by a system of coupled differential equations), and then must benefit from a rare stochastic fluctuation to find the ground state. In the case of WalkSAT we found the fluctuation rate --and hence the time to find a solution-- with a simple Brownian motion (and a corresponding Fokker-Planck) analysis of the algorithm. This was possible exactly because the algorithm's equilibrium state is at high energy, so the region where strong correlations exists between the variables is not explored. 

We have also introduced a new class of quasi-greedy algorithms which are not able to jump over large energy barrier. We have shown that algorithms belonging to this class are very effective in reaching the threshold energy in times $O(N)$ and also in reaching the ground state by rare fluctuations in times $O(\exp(\mu N))$, with an optimal value for the $\mu$ exponent. Given that these algorithms are not able to jump over large energy barriers, they provide a direct evidence that the most effective search for solutions in the hard XORSAT problems is limited mainly by entropic barriers. As it is likely to happen in many other hard optimization problems.

Very recently, an algorithm from this class of quasi-greedy searches has been used in a challenge, whose participants were asked to find the ground state of a 3-regular 3-XORSAT problem \cite{USCchallenge}. At present a very optimized version of this quasi-greedy algorithm, running on Nvidia VT100 GPU, is leading the challenge \cite{leadingXORSAT}, supporting our claim that the quasi-greedy algorithm introduced here are very effective, as the real barrier is entropic in nature.

We have shown, by a time integration of the time-dependent Schrödinger equation, that a similar situation is encountered by quantum annealing algorithms. If one posits to find the solution of the problem with an $O(1)$ probability as $N$ grows, the same rate of growth of the solution time is found for problems with one solution (for which one can apply the adiabatic theorem connecting the gap with the time, as in previous studies \cite{young2010first}) and those with many solutions, for which most of the action occurs at finite energy density, where entropy dominates. Here, in particular, it is not difficult to make the parallel with the difficulties attributed to localization phenomena \cite{altshuler2010anderson}, where many small avoided crossings occur between states which are $O(N)$ spin flip apart.

Therefore we conclude that, in situations in which algorithms work in regions of the phase space in which trapping in local minima is not a real problem, the real problem is entropic barriers, which get for themselves the task of making the solution time exponential. The fact that two such phenomena, apparently so different from each other, can trade places and make sure P$\neq$NP is a fascinating, and we believe not widely appreciated aspect of complexity theory. That quantum algorithms might suffer from a similar trade-off is, if possible, even more surprising.

\begin{acknowledgments}
This work was supported by the European Research Council under the European Union's Horizon 2020 research and innovation program (Grant No.~694925-Lotglassy). We warmly thank Guilhem Semerijan for the careful reading and suggestions.
\end{acknowledgments}

\appendix

\section{Counting blocked configurations and the size of their basins of attraction}
\label{app:countingBlocked}

The greedy version of the algorithm we have introduced has many possible stopping configurations. Indeed any configuration where each variable participate to at most one violated interaction is a local minimum of the energy function and thus can block the greedy dynamics.
We call these configurations \emph{blocked}, because by seeing the dynamics as the evolution of energy defects that can just annihilate in pairs, in a blocked configuration energy defects cannot evolve since they are isolated.

\begin{figure}
    \includegraphics[width=\columnwidth]{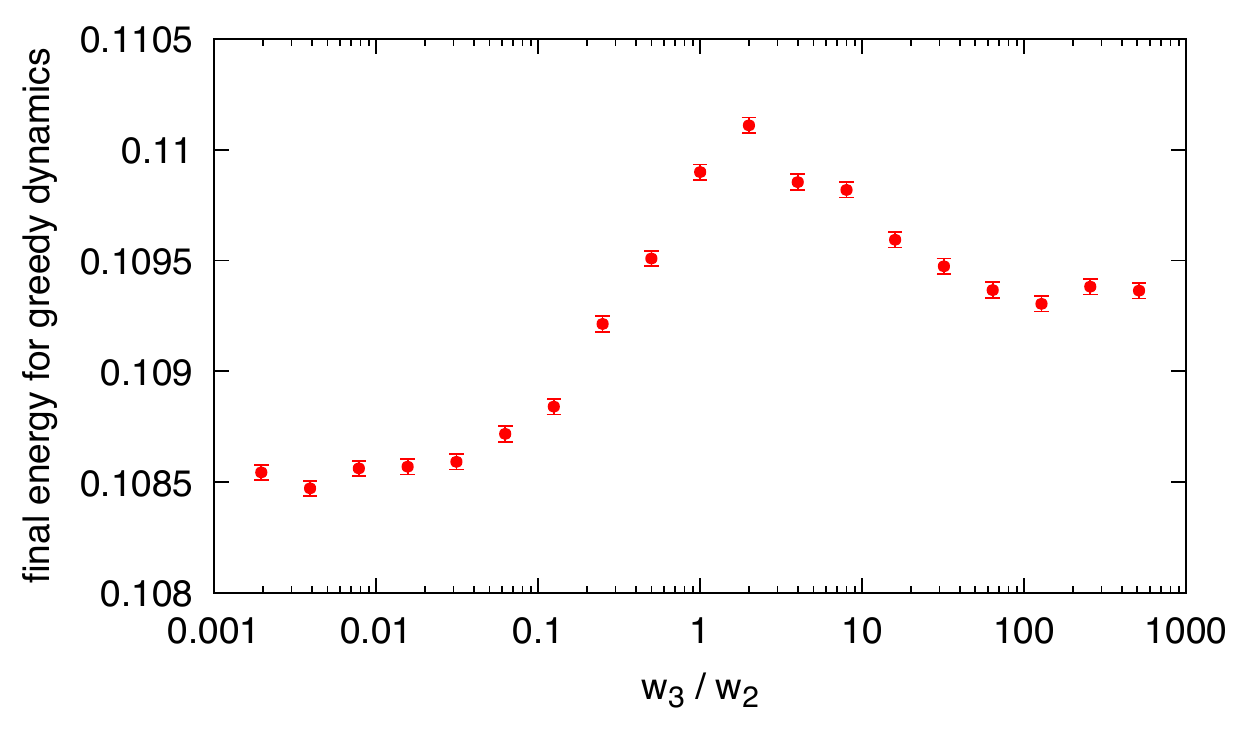}
    \caption{Mean energy of the stopping configurations for the greedy dynamics as a function of the ratio $w_3/w_2$.}
    \label{fig:greedyThresholds}
\end{figure}

We start from the observation that all greedy dynamics tend to a threshold energy around 0.11; more precisely in Fig.~\ref{fig:greedyThresholds} we plot the energy of the stopping configurations for several greedy dynamics as a function of the ratio $w_3 / w_2$ (the only degree of freedom of the algorithm once we fix $w_0=w_1=0$).

It is thus natural to ask whether it is possible to relate the stopping energy of the greedy dynamics to the entropy of blocked configurations.
In order to compute the latter we use the replica symmetric cavity method, that is the Bethe approximation.

We consider the dual lattice to a 3-regular random 3-hypergraph, which is again a 3-regular random 3-hypergraph, where the variables are now the constraints and we assign variables $n_i\in\{0,1\}$ indicating whether a constraint is satisfied ($n_i=0$) or not ($n_i=1$). The interaction among triplets of constraints (those shared by a variable in the original model) forbids any configuration with more than one violated constraint per triplet ($n_1+n_2+n_3\le 1$).

\begin{figure}
\includegraphics[width=0.6\columnwidth]{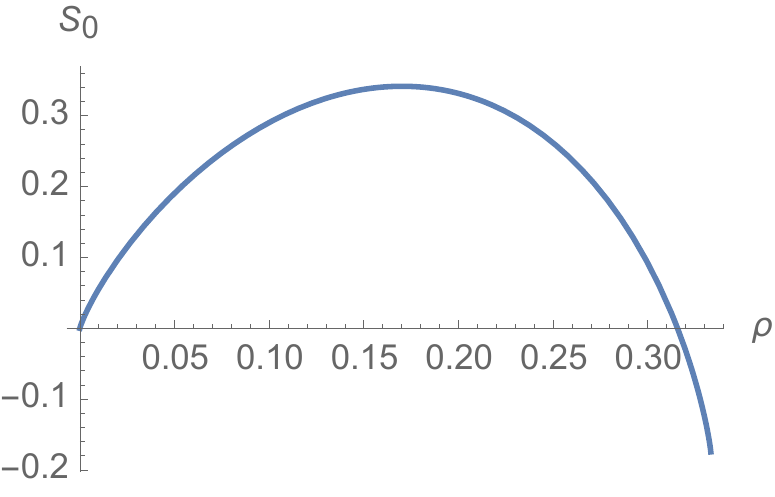}
    \caption{Entropy $S_0$ of blocked configurations as a function of the energy $\rho$.}
    \label{fig:S0}
\end{figure}

The energy of a spin configuration now corresponds to the density of variables $n_i=1$ that we call $\rho$. Under the Bethe approximation the joint probability distribution can be factorized as follows
\begin{equation}
P(\{n_i\}) \simeq \prod_{i=1}^N p_1(n_i) \prod_{(ijk)}^N \frac{p_3(n_i, n_j, n_k)}{p_1(n_1) p_1(n_j) p_1(n_k)}
\label{eq:betheMeas}
\end{equation}
where the second product is over the $N$ randomly chosen triplets forming the 3-regular 3-hypergraph. Given that the random hypergraph is regular we can assume the one-particle and 3-particles marginal probabilities, $p_1(n_i)$ and $p_3(n_i,n_j,n_k)$, being site independent.
These can be written explicitly in terms of the energy $\rho$ as 
\begin{equation}
    p_1(n) = \left\{
    \begin{array}{ll}
         1-\rho &  \text{ if } n=0\\
         \rho & \text{ if } n=1
    \end{array}
    \right.
\end{equation}
\begin{equation}
    p_3(n_1,n_2,n_3) = \left\{
    \begin{array}{ll}
         1-3\rho &  \text{ if } n_1+n_2+n_3=0\\
         \rho & \text{ if } n_1+n_2+n_3=1
    \end{array}
    \right.
\end{equation}
From the Bethe approximation in Eq.~(\ref{eq:betheMeas}) the entropy of blocked configurations can be written as
\begin{multline}
S_0 = - \sum_{n_1,n_2,n_3} p_3(n_1,n_2,n_3) \log p_3(n_1,n_2,n_3) +\\
2\sum_n p_1(n) \log p_1(n) =\\
-(1-3\rho)\log(1-3\rho)-\rho\log\rho+2(1-\rho)\log(1-\rho)
\end{multline}
which is plotted in Fig.~\ref{fig:S0}.

$S_0$ is non negative for $\rho\le 0.315742$ and has a maximum in $\rho^\star = 0.170209$.
The threshold energy for the greedy algorithms we have studied is in the range $[0,\rho^\star]$, but we cannot estimated it from $S_0$ alone, because each blocked configuration has a basin of attraction whose size matters as much as the entropy of blocked configurations.

\begin{figure}
    \centering
    \includegraphics[width=\columnwidth]{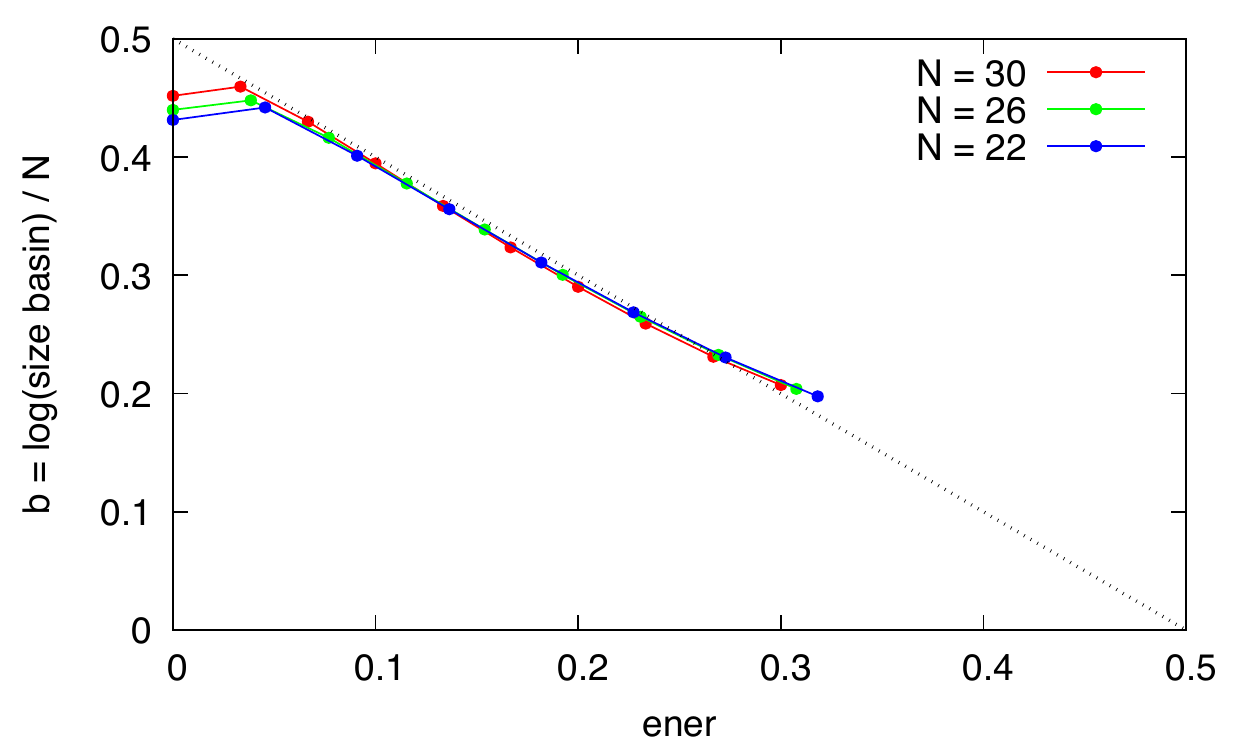}
    \caption{The size of the basin of attraction grows exponentially with the size and the depth of the energy minimum. Finite size effects are evident only for the lowest energies.}
    \label{fig:basins}
\end{figure}

Given the important role of the basins of attraction in predicting the large time limit of relaxation dynamics we have measure their sizes in problems of small size, $N\le30$, where an exact enumeration can be performed. For each blocked configuration we have measured the size of the basin of attraction as the number of initial conditions that a greedy dynamics brings to that blocked configuration. These sizes are in general exponentially large in $N$ and thus we define $b=\log(\text{size of basin})/N$.

We report in Fig.~\ref{fig:basins} the results for $b$ as a function of the energy of the blocked configuration for different sizes. The data have been averaged over different samples and different blocked configuration at fixed energy.
It is remarkable that even for such small sizes the data show rather weak size dependence and are thus reliable.

\begin{figure}
    \centering
    \includegraphics[width=\columnwidth]{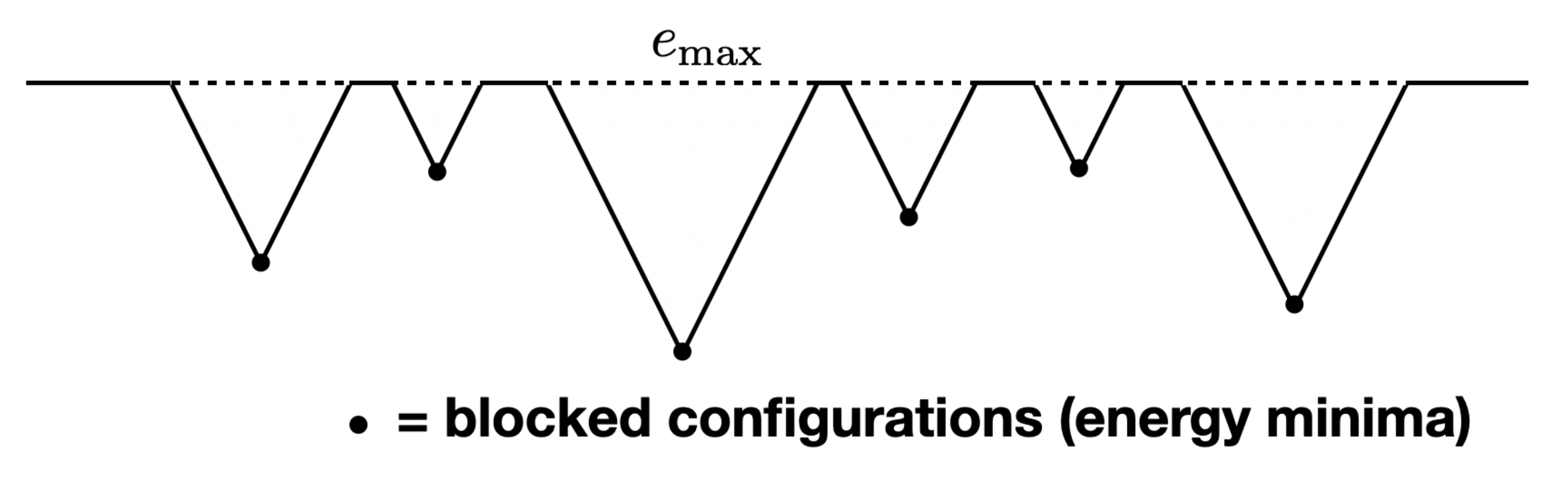}
    \caption{A schematic picture of the energy landscape in the 3-regular 3-XORSAT problem. Most energy minima (including the solutions to the problem) have a basin of attraction which is likely to be connected to the energy level $\emax$ where the dynamics starts. However the entropic barrier makes very unlikely to choose the pit leading to a solution.}
    \label{fig:schematic}
\end{figure}

The dotted line in the figure is just a guide for the eyes to convince the reader that the observed $b(e)$ is not far from following a linear behavior up to the energy of most numerous configurations $\emax=1/2$.
According to this linear behavior the size of a basin of attraction depends linearly on the depth of the energy minimum (the blocked configuration). The simplest picture compatible with these data is the one schematically represented in Fig.~\ref{fig:schematic} where each energy minimum corresponds to a pit whose edge is close to $\emax$, such that the size of the pit grows exponentially with its depth, that is the distance from the edge.

According to the simplified picture in Fig.~\ref{fig:schematic} every pit is in principle accessible from the initial configuration (which has typically an energy $\emax$). However those leading to a solution correspond to a tiny minority of the configurations at $\emax$ and thus is extremely unlikely that the dynamics enter one of these (this is essentially the origin of the entropic barrier).

\begin{figure}
    \includegraphics[width=\columnwidth]{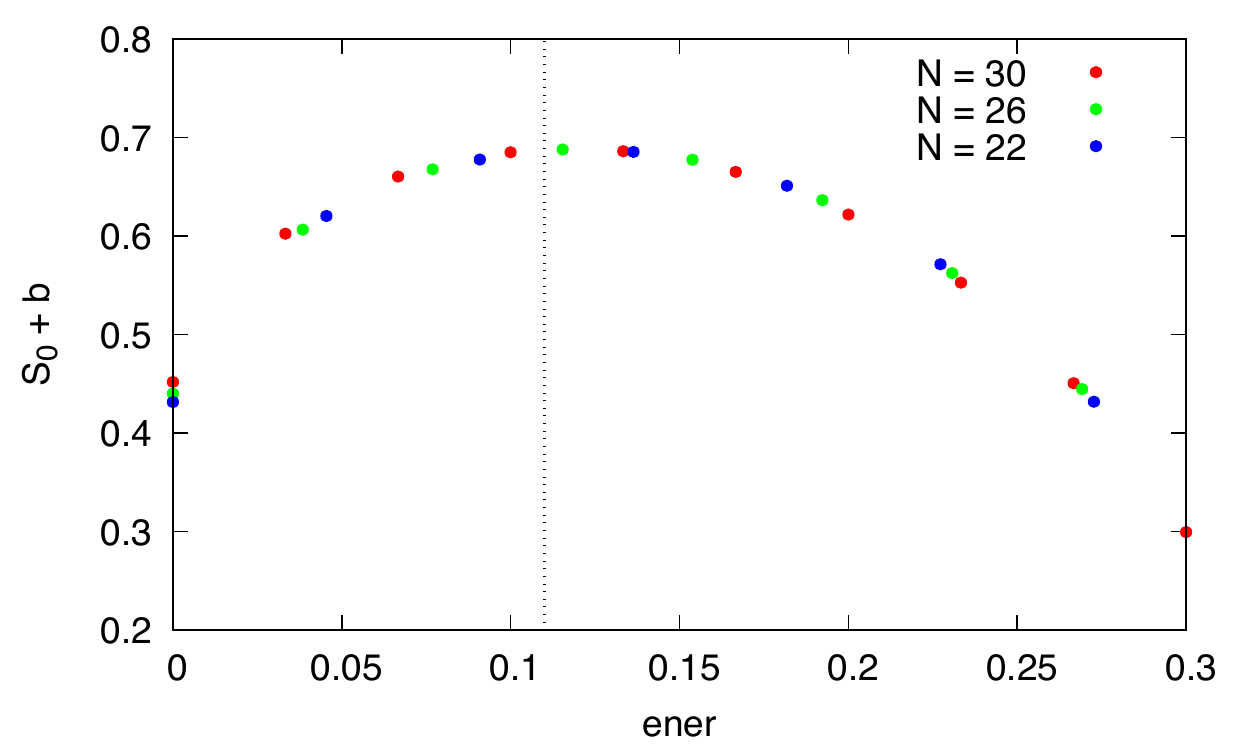}
    \caption{Combining the entropy of blocked configurations $S_0$ and the log size of the basins of attractions $b$ we can compute the rate of the large deviation probability of finding a block configuration of a given energy (shifted by $\log(2)$ because it is not normalized). The maximum is achieved at an energy close to 0.11 (marked by a vertical dotted line) where the greedy algorithm converges in the large $N$ limit.}
    \label{fig:basinsPlusEntro}
\end{figure}

Combining the number of energy minima $\exp(N S_0)$ with their size $\exp(N b)$ we can compute the large deviation rate to find one of the energy minima as a function of their energy. This is shown in Fig.~\ref{fig:basinsPlusEntro}, where indeed the maximum is achieved at $e \sim 0.11$ (marked by a vertical dotted line) which is the typical energy reached by the greedy algorithm in the large $N$ limit.
For finite values of $N$ these data also provide the exponential rate to find a solution via the greedy algorithm: this is just the probability of randomly choosing one of the pits having the bottom at $e=0$ and corresponds to $\mu=\max_e(S_0+b) - (S_0+b)|_{e=0} \approx 0.2$.

\section{An optimal temperature scheduling for Parallel Tempering}
\label{app:PTscheduling}

The main problem in setting up an optimized PT is the choice of the temperature set. However in the present model we are in a lucky situation because we known that the 3-regular random 3-XORSAT has no thermodynamical phase transition and so, after convergence, the energy sampled by the PT must be the one of the paramagnetic solution, $e(\beta)=[1-\tanh(\beta/2)]/2$.
In the large $N$ limit we can assume that the extensive energy at inverse temperature $\beta$ is a Gaussian variables with mean $E(\beta)=N e(\beta)$ and variance $\sigma^2(\beta)=-N e'(\beta)$. This Gaussianity assumption (which is rather well satisfied, but in the vicinity of the ground state) allows us to compute the probability of swapping two replicas at temperatures $\beta_1$ and $\beta_2$, which has to be a function of the ratio of the energy difference to the energy fluctuation
\[
p_\text{swap} = f\left(\frac{E(\beta_2)-E(\beta_1)}{\sqrt{\sigma^2(\beta_1)+\sigma^2(\beta_2)}}\right)\;.
\]
In the large $N$ limit such a ratio can be written as
\[
\frac{E(\beta_2)-E(\beta_1)}{\sqrt{\sigma^2(\beta_1)+\sigma^2(\beta_2)}} \simeq \frac{N e'(\beta) \Delta\beta}{\sqrt{N e'(\beta)}} \simeq \Delta\beta\sqrt{N e'(\beta)}\;.
\]
The explicit form for the function $f$ is given by
\[
f(z) = \int dx\,dy\,\frac{e^{-\frac{x^2}{2}-\frac{(y-z)^2}{2}}}{2\pi} \min\left(1,e^{z(x-y)}\right)\;.
\]
The best way to allow replicas to wander fast between temperatures is to fix a constant $p_\text{swap}$ between any pair of successive temperatures, that is using temperature intervals set by
\[
\Delta \beta = \frac{f^{-1}(p_\text{swap})}{\sqrt{N e'(\beta)}}\;,
\]
which implies in the large $N$ limit the following recursive equation for the temperatures to be used in the optimized Parallel Tempering
\[
\beta_{n+1} = \beta_n + \frac{2r}{\sqrt{N(1-\tanh(\beta_n/2)^2)}}
\]
where $r=f^{-1}(p_\text{swap})$. The solution to the above recursive equation converges in the large $N$ limit to the following set of temperatures
\begin{equation}
\beta_n = 2 \;\text{arcsinh}\left[\tan\left(\frac{r n}{\sqrt{N}}\right)\right]\;,
\label{eq:betasPT}
\end{equation}
with the index $n$ running up to $\lfloor \sqrt{N}\pi/(2r)\rfloor$.

The optimal value for $r$ is the one minimizing the mean traveling time between the extremal temperatures, which is proportional to $[r^2 f(r)]^{-1}$. The minimum is achieved at $r_\text{opt} \approx 1.68$, that corresponds to an optimal swapping rate $f(r_\text{opt})\approx 0.23$ (this result is well known as the 0.23 rule). We run all out PT simulations with the temperature set defined in Eq.~(\ref{eq:betasPT}) with $r=r_\text{opt}$.

\bibliography{biblio.bib}

\begin{thebibliography}{63}
\expandafter\ifx\csname natexlab\endcsname\relax\def\natexlab#1{#1}\fi
\expandafter\ifx\csname bibnamefont\endcsname\relax
  \def\bibnamefont#1{#1}\fi
\expandafter\ifx\csname bibfnamefont\endcsname\relax
  \def\bibfnamefont#1{#1}\fi
\expandafter\ifx\csname citenamefont\endcsname\relax
  \def\citenamefont#1{#1}\fi
\expandafter\ifx\csname url\endcsname\relax
  \def\url#1{\texttt{#1}}\fi
\expandafter\ifx\csname urlprefix\endcsname\relax\def\urlprefix{URL }\fi
\providecommand{\bibinfo}[2]{#2}
\providecommand{\eprint}[2][]{\url{#2}}

\bibitem[{\citenamefont{Arora and Barak}(2009)}]{arora2009computational}
\bibinfo{author}{\bibfnamefont{S.}~\bibnamefont{Arora}} \bibnamefont{and}
  \bibinfo{author}{\bibfnamefont{B.}~\bibnamefont{Barak}},
  \emph{\bibinfo{title}{Computational complexity: a modern approach}}
  (\bibinfo{publisher}{Cambridge University Press}, \bibinfo{year}{2009}).

\bibitem[{\citenamefont{Hartmann and Weigt}(2005)}]{hartmann2005phase}
\bibinfo{author}{\bibfnamefont{A.~K.} \bibnamefont{Hartmann}} \bibnamefont{and}
  \bibinfo{author}{\bibfnamefont{M.}~\bibnamefont{Weigt}},
  \emph{\bibinfo{title}{Phase transitions in combinatorial optimization
  problems}}, vol.~\bibinfo{volume}{67} (\bibinfo{publisher}{Wiley Online
  Library}, \bibinfo{year}{2005}).

\bibitem[{\citenamefont{Ricci-Tersenghi
  et~al.}(2001)\citenamefont{Ricci-Tersenghi, Weigt, and
  Zecchina}}]{ricci2001simplest}
\bibinfo{author}{\bibfnamefont{F.}~\bibnamefont{Ricci-Tersenghi}},
  \bibinfo{author}{\bibfnamefont{M.}~\bibnamefont{Weigt}}, \bibnamefont{and}
  \bibinfo{author}{\bibfnamefont{R.}~\bibnamefont{Zecchina}},
  \bibinfo{journal}{Physical Review E} \textbf{\bibinfo{volume}{63}},
  \bibinfo{pages}{026702} (\bibinfo{year}{2001}).

\bibitem[{\citenamefont{M{\'e}zard et~al.}(2002)\citenamefont{M{\'e}zard,
  Parisi, and Zecchina}}]{mezard2002analytic}
\bibinfo{author}{\bibfnamefont{M.}~\bibnamefont{M{\'e}zard}},
  \bibinfo{author}{\bibfnamefont{G.}~\bibnamefont{Parisi}}, \bibnamefont{and}
  \bibinfo{author}{\bibfnamefont{R.}~\bibnamefont{Zecchina}},
  \bibinfo{journal}{Science} \textbf{\bibinfo{volume}{297}},
  \bibinfo{pages}{812} (\bibinfo{year}{2002}).

\bibitem[{\citenamefont{Barthel et~al.}(2002)\citenamefont{Barthel, Hartmann,
  Leone, Ricci-Tersenghi, Weigt, and Zecchina}}]{barthel2002hiding}
\bibinfo{author}{\bibfnamefont{W.}~\bibnamefont{Barthel}},
  \bibinfo{author}{\bibfnamefont{A.~K.} \bibnamefont{Hartmann}},
  \bibinfo{author}{\bibfnamefont{M.}~\bibnamefont{Leone}},
  \bibinfo{author}{\bibfnamefont{F.}~\bibnamefont{Ricci-Tersenghi}},
  \bibinfo{author}{\bibfnamefont{M.}~\bibnamefont{Weigt}}, \bibnamefont{and}
  \bibinfo{author}{\bibfnamefont{R.}~\bibnamefont{Zecchina}},
  \bibinfo{journal}{Physical review letters} \textbf{\bibinfo{volume}{88}},
  \bibinfo{pages}{188701} (\bibinfo{year}{2002}).

\bibitem[{\citenamefont{M{\'e}zard et~al.}(2003)\citenamefont{M{\'e}zard,
  Ricci-Tersenghi, and Zecchina}}]{mezard2003two}
\bibinfo{author}{\bibfnamefont{M.}~\bibnamefont{M{\'e}zard}},
  \bibinfo{author}{\bibfnamefont{F.}~\bibnamefont{Ricci-Tersenghi}},
  \bibnamefont{and} \bibinfo{author}{\bibfnamefont{R.}~\bibnamefont{Zecchina}},
  \bibinfo{journal}{Journal of Statistical Physics}
  \textbf{\bibinfo{volume}{111}}, \bibinfo{pages}{505} (\bibinfo{year}{2003}).

\bibitem[{\citenamefont{Cocco et~al.}(2003{\natexlab{a}})\citenamefont{Cocco,
  Dubois, Mandler, and Monasson}}]{cocco2003rigorous}
\bibinfo{author}{\bibfnamefont{S.}~\bibnamefont{Cocco}},
  \bibinfo{author}{\bibfnamefont{O.}~\bibnamefont{Dubois}},
  \bibinfo{author}{\bibfnamefont{J.}~\bibnamefont{Mandler}}, \bibnamefont{and}
  \bibinfo{author}{\bibfnamefont{R.}~\bibnamefont{Monasson}},
  \bibinfo{journal}{Physical review letters} \textbf{\bibinfo{volume}{90}},
  \bibinfo{pages}{047205} (\bibinfo{year}{2003}{\natexlab{a}}).

\bibitem[{\citenamefont{Montanari et~al.}(2004)\citenamefont{Montanari, Parisi,
  and Ricci-Tersenghi}}]{montanari2004instability}
\bibinfo{author}{\bibfnamefont{A.}~\bibnamefont{Montanari}},
  \bibinfo{author}{\bibfnamefont{G.}~\bibnamefont{Parisi}}, \bibnamefont{and}
  \bibinfo{author}{\bibfnamefont{F.}~\bibnamefont{Ricci-Tersenghi}},
  \bibinfo{journal}{Journal of Physics A: Mathematical and General}
  \textbf{\bibinfo{volume}{37}}, \bibinfo{pages}{2073} (\bibinfo{year}{2004}).

\bibitem[{\citenamefont{M{\'e}zard et~al.}(2005)\citenamefont{M{\'e}zard, Mora,
  and Zecchina}}]{mezard2005clustering}
\bibinfo{author}{\bibfnamefont{M.}~\bibnamefont{M{\'e}zard}},
  \bibinfo{author}{\bibfnamefont{T.}~\bibnamefont{Mora}}, \bibnamefont{and}
  \bibinfo{author}{\bibfnamefont{R.}~\bibnamefont{Zecchina}},
  \bibinfo{journal}{Physical Review Letters} \textbf{\bibinfo{volume}{94}},
  \bibinfo{pages}{197205} (\bibinfo{year}{2005}).

\bibitem[{\citenamefont{Mertens et~al.}(2006)\citenamefont{Mertens, M{\'e}zard,
  and Zecchina}}]{mertens2006threshold}
\bibinfo{author}{\bibfnamefont{S.}~\bibnamefont{Mertens}},
  \bibinfo{author}{\bibfnamefont{M.}~\bibnamefont{M{\'e}zard}},
  \bibnamefont{and} \bibinfo{author}{\bibfnamefont{R.}~\bibnamefont{Zecchina}},
  \bibinfo{journal}{Random Structures \& Algorithms}
  \textbf{\bibinfo{volume}{28}}, \bibinfo{pages}{340} (\bibinfo{year}{2006}).

\bibitem[{\citenamefont{Krzaka{\l}a et~al.}(2007)\citenamefont{Krzaka{\l}a,
  Montanari, Ricci-Tersenghi, Semerjian, and
  Zdeborov{\'a}}}]{krzakala2007gibbs}
\bibinfo{author}{\bibfnamefont{F.}~\bibnamefont{Krzaka{\l}a}},
  \bibinfo{author}{\bibfnamefont{A.}~\bibnamefont{Montanari}},
  \bibinfo{author}{\bibfnamefont{F.}~\bibnamefont{Ricci-Tersenghi}},
  \bibinfo{author}{\bibfnamefont{G.}~\bibnamefont{Semerjian}},
  \bibnamefont{and}
  \bibinfo{author}{\bibfnamefont{L.}~\bibnamefont{Zdeborov{\'a}}},
  \bibinfo{journal}{Proceedings of the National Academy of Sciences}
  \textbf{\bibinfo{volume}{104}}, \bibinfo{pages}{10318}
  (\bibinfo{year}{2007}).

\bibitem[{\citenamefont{Krzakala and Kurchan}(2007)}]{krzakala2007landscape}
\bibinfo{author}{\bibfnamefont{F.}~\bibnamefont{Krzakala}} \bibnamefont{and}
  \bibinfo{author}{\bibfnamefont{J.}~\bibnamefont{Kurchan}},
  \bibinfo{journal}{Physical Review E} \textbf{\bibinfo{volume}{76}},
  \bibinfo{pages}{021122} (\bibinfo{year}{2007}).

\bibitem[{\citenamefont{Montanari et~al.}(2008)\citenamefont{Montanari,
  Ricci-Tersenghi, and Semerjian}}]{montanari2008clusters}
\bibinfo{author}{\bibfnamefont{A.}~\bibnamefont{Montanari}},
  \bibinfo{author}{\bibfnamefont{F.}~\bibnamefont{Ricci-Tersenghi}},
  \bibnamefont{and}
  \bibinfo{author}{\bibfnamefont{G.}~\bibnamefont{Semerjian}},
  \bibinfo{journal}{Journal of Statistical Mechanics: Theory and Experiment}
  \textbf{\bibinfo{volume}{2008}}, \bibinfo{pages}{P04004}
  (\bibinfo{year}{2008}).

\bibitem[{\citenamefont{Zdeborov{\'a} and
  Krzaka{\l}a}(2007)}]{zdeborova2007phase}
\bibinfo{author}{\bibfnamefont{L.}~\bibnamefont{Zdeborov{\'a}}}
  \bibnamefont{and}
  \bibinfo{author}{\bibfnamefont{F.}~\bibnamefont{Krzaka{\l}a}},
  \bibinfo{journal}{Physical Review E} \textbf{\bibinfo{volume}{76}},
  \bibinfo{pages}{031131} (\bibinfo{year}{2007}).

\bibitem[{\citenamefont{Altarelli et~al.}(2008)\citenamefont{Altarelli,
  Monasson, and Zamponi}}]{altarelli2008relationship}
\bibinfo{author}{\bibfnamefont{F.}~\bibnamefont{Altarelli}},
  \bibinfo{author}{\bibfnamefont{R.}~\bibnamefont{Monasson}}, \bibnamefont{and}
  \bibinfo{author}{\bibfnamefont{F.}~\bibnamefont{Zamponi}}, in
  \emph{\bibinfo{booktitle}{Journal of Physics: Conference Series}}
  (\bibinfo{year}{2008}), vol.~\bibinfo{volume}{95}, p.
  \bibinfo{pages}{012013}.

\bibitem[{\citenamefont{Krzakala et~al.}(2016)\citenamefont{Krzakala,
  Ricci-Tersenghi, Zdeborova, Zecchina, Tramel, and
  Cugliandolo}}]{krzakala2016statistical}
\bibinfo{author}{\bibfnamefont{F.}~\bibnamefont{Krzakala}},
  \bibinfo{author}{\bibfnamefont{F.}~\bibnamefont{Ricci-Tersenghi}},
  \bibinfo{author}{\bibfnamefont{L.}~\bibnamefont{Zdeborova}},
  \bibinfo{author}{\bibfnamefont{R.}~\bibnamefont{Zecchina}},
  \bibinfo{author}{\bibfnamefont{E.~W.} \bibnamefont{Tramel}},
  \bibnamefont{and} \bibinfo{author}{\bibfnamefont{L.~F.}
  \bibnamefont{Cugliandolo}}, \emph{\bibinfo{title}{Statistical Physics,
  Optimization, Inference, and Message-Passing Algorithms: Lecture Notes of the
  Les Houches School of Physics-Special Issue, October 2013}},
  \bibinfo{number}{2013} (\bibinfo{publisher}{Oxford University Press},
  \bibinfo{year}{2016}).

\bibitem[{\citenamefont{Farhi et~al.}(2000)\citenamefont{Farhi, Goldstone,
  Gutmann, and Sipser}}]{farhi2000quantum}
\bibinfo{author}{\bibfnamefont{E.}~\bibnamefont{Farhi}},
  \bibinfo{author}{\bibfnamefont{J.}~\bibnamefont{Goldstone}},
  \bibinfo{author}{\bibfnamefont{S.}~\bibnamefont{Gutmann}}, \bibnamefont{and}
  \bibinfo{author}{\bibfnamefont{M.}~\bibnamefont{Sipser}},
  \bibinfo{journal}{arXiv preprint quant-ph/0001106}  (\bibinfo{year}{2000}).

\bibitem[{\citenamefont{Santoro and Tosatti}(2006)}]{santoro2006optimization}
\bibinfo{author}{\bibfnamefont{G.~E.} \bibnamefont{Santoro}} \bibnamefont{and}
  \bibinfo{author}{\bibfnamefont{E.}~\bibnamefont{Tosatti}},
  \bibinfo{journal}{Journal of Physics A: Mathematical and General}
  \textbf{\bibinfo{volume}{39}}, \bibinfo{pages}{R393} (\bibinfo{year}{2006}).

\bibitem[{\citenamefont{Farhi et~al.}(2012)\citenamefont{Farhi, Gosset, Hen,
  Sandvik, Shor, Young, and Zamponi}}]{farhi2012performance}
\bibinfo{author}{\bibfnamefont{E.}~\bibnamefont{Farhi}},
  \bibinfo{author}{\bibfnamefont{D.}~\bibnamefont{Gosset}},
  \bibinfo{author}{\bibfnamefont{I.}~\bibnamefont{Hen}},
  \bibinfo{author}{\bibfnamefont{A.}~\bibnamefont{Sandvik}},
  \bibinfo{author}{\bibfnamefont{P.}~\bibnamefont{Shor}},
  \bibinfo{author}{\bibfnamefont{A.}~\bibnamefont{Young}}, \bibnamefont{and}
  \bibinfo{author}{\bibfnamefont{F.}~\bibnamefont{Zamponi}},
  \bibinfo{journal}{Physical Review A} \textbf{\bibinfo{volume}{86}},
  \bibinfo{pages}{052334} (\bibinfo{year}{2012}).

\bibitem[{\citenamefont{Altshuler et~al.}(2010)\citenamefont{Altshuler, Krovi,
  and Roland}}]{altshuler2010anderson}
\bibinfo{author}{\bibfnamefont{B.}~\bibnamefont{Altshuler}},
  \bibinfo{author}{\bibfnamefont{H.}~\bibnamefont{Krovi}}, \bibnamefont{and}
  \bibinfo{author}{\bibfnamefont{J.}~\bibnamefont{Roland}},
  \bibinfo{journal}{Proceedings of the National Academy of Sciences}
  \textbf{\bibinfo{volume}{107}}, \bibinfo{pages}{12446}
  (\bibinfo{year}{2010}).

\bibitem[{\citenamefont{Farhi et~al.}(2014)\citenamefont{Farhi, Goldstone, and
  Gutmann}}]{farhi2014quantum}
\bibinfo{author}{\bibfnamefont{E.}~\bibnamefont{Farhi}},
  \bibinfo{author}{\bibfnamefont{J.}~\bibnamefont{Goldstone}},
  \bibnamefont{and} \bibinfo{author}{\bibfnamefont{S.}~\bibnamefont{Gutmann}},
  \bibinfo{journal}{arXiv preprint arXiv:1411.4028}  (\bibinfo{year}{2014}).

\bibitem[{\citenamefont{Laumann et~al.}(2015)\citenamefont{Laumann, Moessner,
  Scardicchio, and Sondhi}}]{laumann2015quantum}
\bibinfo{author}{\bibfnamefont{C.~R.} \bibnamefont{Laumann}},
  \bibinfo{author}{\bibfnamefont{R.}~\bibnamefont{Moessner}},
  \bibinfo{author}{\bibfnamefont{A.}~\bibnamefont{Scardicchio}},
  \bibnamefont{and} \bibinfo{author}{\bibfnamefont{S.~L.}
  \bibnamefont{Sondhi}}, \bibinfo{journal}{The European Physical Journal
  Special Topics} \textbf{\bibinfo{volume}{224}}, \bibinfo{pages}{75}
  (\bibinfo{year}{2015}).

\bibitem[{\citenamefont{Mossi and Scardicchio}(2017)}]{mossi2017ergodic}
\bibinfo{author}{\bibfnamefont{G.}~\bibnamefont{Mossi}} \bibnamefont{and}
  \bibinfo{author}{\bibfnamefont{A.}~\bibnamefont{Scardicchio}},
  \bibinfo{journal}{Philosophical Transactions of the Royal Society A:
  Mathematical, Physical and Engineering Sciences}
  \textbf{\bibinfo{volume}{375}}, \bibinfo{pages}{20160424}
  (\bibinfo{year}{2017}).

\bibitem[{\citenamefont{Mossi et~al.}(2017)}]{mossi2017transverse}
\bibinfo{author}{\bibfnamefont{G.}~\bibnamefont{Mossi}} \bibnamefont{et~al.}
  (\bibinfo{year}{2017}).

\bibitem[{\citenamefont{Smelyanskiy et~al.}(2020)\citenamefont{Smelyanskiy,
  Kechedzhi, Boixo, Isakov, Neven, and Altshuler}}]{smelyanskiy2020nonergodic}
\bibinfo{author}{\bibfnamefont{V.~N.} \bibnamefont{Smelyanskiy}},
  \bibinfo{author}{\bibfnamefont{K.}~\bibnamefont{Kechedzhi}},
  \bibinfo{author}{\bibfnamefont{S.}~\bibnamefont{Boixo}},
  \bibinfo{author}{\bibfnamefont{S.~V.} \bibnamefont{Isakov}},
  \bibinfo{author}{\bibfnamefont{H.}~\bibnamefont{Neven}}, \bibnamefont{and}
  \bibinfo{author}{\bibfnamefont{B.}~\bibnamefont{Altshuler}},
  \bibinfo{journal}{Physical Review X} \textbf{\bibinfo{volume}{10}},
  \bibinfo{pages}{011017} (\bibinfo{year}{2020}).

\bibitem[{\citenamefont{Basko et~al.}(2006)\citenamefont{Basko, Aleiner, and
  Altshuler}}]{basko2006metal}
\bibinfo{author}{\bibfnamefont{D.~M.} \bibnamefont{Basko}},
  \bibinfo{author}{\bibfnamefont{I.~L.} \bibnamefont{Aleiner}},
  \bibnamefont{and} \bibinfo{author}{\bibfnamefont{B.~L.}
  \bibnamefont{Altshuler}}, \bibinfo{journal}{Annals of physics}
  \textbf{\bibinfo{volume}{321}}, \bibinfo{pages}{1126} (\bibinfo{year}{2006}).

\bibitem[{\citenamefont{Laumann et~al.}(2014)\citenamefont{Laumann, Pal, and
  Scardicchio}}]{laumann2014many}
\bibinfo{author}{\bibfnamefont{C.~R.} \bibnamefont{Laumann}},
  \bibinfo{author}{\bibfnamefont{A.}~\bibnamefont{Pal}}, \bibnamefont{and}
  \bibinfo{author}{\bibfnamefont{A.}~\bibnamefont{Scardicchio}},
  \bibinfo{journal}{Physical review letters} \textbf{\bibinfo{volume}{113}},
  \bibinfo{pages}{200405} (\bibinfo{year}{2014}).

\bibitem[{\citenamefont{Imbrie et~al.}(2017)\citenamefont{Imbrie, Ros, and
  Scardicchio}}]{imbrie2017local}
\bibinfo{author}{\bibfnamefont{J.~Z.} \bibnamefont{Imbrie}},
  \bibinfo{author}{\bibfnamefont{V.}~\bibnamefont{Ros}}, \bibnamefont{and}
  \bibinfo{author}{\bibfnamefont{A.}~\bibnamefont{Scardicchio}},
  \bibinfo{journal}{Annalen der Physik} \textbf{\bibinfo{volume}{529}},
  \bibinfo{pages}{1600278} (\bibinfo{year}{2017}).

\bibitem[{\citenamefont{Arute et~al.}(2019)\citenamefont{Arute, Arya, Babbush,
  Bacon, Bardin, Barends, Biswas, Boixo, Brandao, Buell
  et~al.}}]{arute2019quantum}
\bibinfo{author}{\bibfnamefont{F.}~\bibnamefont{Arute}},
  \bibinfo{author}{\bibfnamefont{K.}~\bibnamefont{Arya}},
  \bibinfo{author}{\bibfnamefont{R.}~\bibnamefont{Babbush}},
  \bibinfo{author}{\bibfnamefont{D.}~\bibnamefont{Bacon}},
  \bibinfo{author}{\bibfnamefont{J.~C.} \bibnamefont{Bardin}},
  \bibinfo{author}{\bibfnamefont{R.}~\bibnamefont{Barends}},
  \bibinfo{author}{\bibfnamefont{R.}~\bibnamefont{Biswas}},
  \bibinfo{author}{\bibfnamefont{S.}~\bibnamefont{Boixo}},
  \bibinfo{author}{\bibfnamefont{F.~G.} \bibnamefont{Brandao}},
  \bibinfo{author}{\bibfnamefont{D.~A.} \bibnamefont{Buell}},
  \bibnamefont{et~al.}, \bibinfo{journal}{Nature}
  \textbf{\bibinfo{volume}{574}}, \bibinfo{pages}{505} (\bibinfo{year}{2019}).

\bibitem[{\citenamefont{Dubois and Mandler}(2002)}]{dubois20023}
\bibinfo{author}{\bibfnamefont{O.}~\bibnamefont{Dubois}} \bibnamefont{and}
  \bibinfo{author}{\bibfnamefont{J.}~\bibnamefont{Mandler}},
  \bibinfo{journal}{Comptes Rendus Mathematique}
  \textbf{\bibinfo{volume}{335}}, \bibinfo{pages}{963} (\bibinfo{year}{2002}).

\bibitem[{\citenamefont{Cocco et~al.}(2003{\natexlab{b}})\citenamefont{Cocco,
  Monasson, Montanari, and Semerjian}}]{cocco2003approximate}
\bibinfo{author}{\bibfnamefont{S.}~\bibnamefont{Cocco}},
  \bibinfo{author}{\bibfnamefont{R.}~\bibnamefont{Monasson}},
  \bibinfo{author}{\bibfnamefont{A.}~\bibnamefont{Montanari}},
  \bibnamefont{and}
  \bibinfo{author}{\bibfnamefont{G.}~\bibnamefont{Semerjian}},
  \bibinfo{journal}{arXiv preprint cs/0302003}
  (\bibinfo{year}{2003}{\natexlab{b}}).

\bibitem[{\citenamefont{Ibrahimi et~al.}(2012)\citenamefont{Ibrahimi, Kanoria,
  Kraning, and Montanari}}]{ibrahimi2012set}
\bibinfo{author}{\bibfnamefont{M.}~\bibnamefont{Ibrahimi}},
  \bibinfo{author}{\bibfnamefont{Y.}~\bibnamefont{Kanoria}},
  \bibinfo{author}{\bibfnamefont{M.}~\bibnamefont{Kraning}}, \bibnamefont{and}
  \bibinfo{author}{\bibfnamefont{A.}~\bibnamefont{Montanari}}, in
  \emph{\bibinfo{booktitle}{Proceedings of the twenty-third annual ACM-SIAM
  symposium on Discrete Algorithms}} (\bibinfo{organization}{SIAM},
  \bibinfo{year}{2012}), pp. \bibinfo{pages}{760--779}.

\bibitem[{\citenamefont{Gamarnik and Jagannath}(2019)}]{gamarnik2019overlap}
\bibinfo{author}{\bibfnamefont{D.}~\bibnamefont{Gamarnik}} \bibnamefont{and}
  \bibinfo{author}{\bibfnamefont{A.}~\bibnamefont{Jagannath}},
  \bibinfo{journal}{arXiv preprint arXiv:1911.06943}  (\bibinfo{year}{2019}).

\bibitem[{\citenamefont{Braunstein et~al.}(2002)\citenamefont{Braunstein,
  Leone, Ricci-Tersenghi, and Zecchina}}]{braunstein2002complexity}
\bibinfo{author}{\bibfnamefont{A.}~\bibnamefont{Braunstein}},
  \bibinfo{author}{\bibfnamefont{M.}~\bibnamefont{Leone}},
  \bibinfo{author}{\bibfnamefont{F.}~\bibnamefont{Ricci-Tersenghi}},
  \bibnamefont{and} \bibinfo{author}{\bibfnamefont{R.}~\bibnamefont{Zecchina}},
  \bibinfo{journal}{Journal of Physics A: Mathematical and General}
  \textbf{\bibinfo{volume}{35}}, \bibinfo{pages}{7559} (\bibinfo{year}{2002}).

\bibitem[{\citenamefont{Franz et~al.}(2001{\natexlab{a}})\citenamefont{Franz,
  M{\'e}zard, Ricci-Tersenghi, Weigt, and Zecchina}}]{franz2001ferromagnet}
\bibinfo{author}{\bibfnamefont{S.}~\bibnamefont{Franz}},
  \bibinfo{author}{\bibfnamefont{M.}~\bibnamefont{M{\'e}zard}},
  \bibinfo{author}{\bibfnamefont{F.}~\bibnamefont{Ricci-Tersenghi}},
  \bibinfo{author}{\bibfnamefont{M.}~\bibnamefont{Weigt}}, \bibnamefont{and}
  \bibinfo{author}{\bibfnamefont{R.}~\bibnamefont{Zecchina}},
  \bibinfo{journal}{EPL (Europhysics Letters)} \textbf{\bibinfo{volume}{55}},
  \bibinfo{pages}{465} (\bibinfo{year}{2001}{\natexlab{a}}).

\bibitem[{\citenamefont{Krzakala and
  Zdeborov{\'a}}(2010)}]{krzakala2010following}
\bibinfo{author}{\bibfnamefont{F.}~\bibnamefont{Krzakala}} \bibnamefont{and}
  \bibinfo{author}{\bibfnamefont{L.}~\bibnamefont{Zdeborov{\'a}}},
  \bibinfo{journal}{EPL (Europhysics Letters)} \textbf{\bibinfo{volume}{90}},
  \bibinfo{pages}{66002} (\bibinfo{year}{2010}).

\bibitem[{\citenamefont{Zdeborov{\'a} and
  Krzakala}(2010)}]{zdeborova2010generalization}
\bibinfo{author}{\bibfnamefont{L.}~\bibnamefont{Zdeborov{\'a}}}
  \bibnamefont{and} \bibinfo{author}{\bibfnamefont{F.}~\bibnamefont{Krzakala}},
  \bibinfo{journal}{Physical Review B} \textbf{\bibinfo{volume}{81}},
  \bibinfo{pages}{224205} (\bibinfo{year}{2010}).

\bibitem[{\citenamefont{Zdeborov{\'a} and
  Krzakala}(2016)}]{zdeborova2016statistical}
\bibinfo{author}{\bibfnamefont{L.}~\bibnamefont{Zdeborov{\'a}}}
  \bibnamefont{and} \bibinfo{author}{\bibfnamefont{F.}~\bibnamefont{Krzakala}},
  \bibinfo{journal}{Advances in Physics} \textbf{\bibinfo{volume}{65}},
  \bibinfo{pages}{453} (\bibinfo{year}{2016}).

\bibitem[{\citenamefont{Franz et~al.}(2001{\natexlab{b}})\citenamefont{Franz,
  Leone, Ricci-Tersenghi, and Zecchina}}]{franz2001exact}
\bibinfo{author}{\bibfnamefont{S.}~\bibnamefont{Franz}},
  \bibinfo{author}{\bibfnamefont{M.}~\bibnamefont{Leone}},
  \bibinfo{author}{\bibfnamefont{F.}~\bibnamefont{Ricci-Tersenghi}},
  \bibnamefont{and} \bibinfo{author}{\bibfnamefont{R.}~\bibnamefont{Zecchina}},
  \bibinfo{journal}{Physical Review Letters} \textbf{\bibinfo{volume}{87}},
  \bibinfo{pages}{127209} (\bibinfo{year}{2001}{\natexlab{b}}).

\bibitem[{\citenamefont{Montanari and
  Ricci-Tersenghi}(2003)}]{montanari2003nature}
\bibinfo{author}{\bibfnamefont{A.}~\bibnamefont{Montanari}} \bibnamefont{and}
  \bibinfo{author}{\bibfnamefont{F.}~\bibnamefont{Ricci-Tersenghi}},
  \bibinfo{journal}{The European Physical Journal B-Condensed Matter and
  Complex Systems} \textbf{\bibinfo{volume}{33}}, \bibinfo{pages}{339}
  (\bibinfo{year}{2003}).

\bibitem[{\citenamefont{Montanari and
  Ricci-Tersenghi}(2004)}]{montanari2004cooling}
\bibinfo{author}{\bibfnamefont{A.}~\bibnamefont{Montanari}} \bibnamefont{and}
  \bibinfo{author}{\bibfnamefont{F.}~\bibnamefont{Ricci-Tersenghi}},
  \bibinfo{journal}{Physical Review B} \textbf{\bibinfo{volume}{70}},
  \bibinfo{pages}{134406} (\bibinfo{year}{2004}).

\bibitem[{\citenamefont{J{\"o}rg et~al.}(2010)\citenamefont{J{\"o}rg, Krzakala,
  Semerjian, and Zamponi}}]{jorg2010first}
\bibinfo{author}{\bibfnamefont{T.}~\bibnamefont{J{\"o}rg}},
  \bibinfo{author}{\bibfnamefont{F.}~\bibnamefont{Krzakala}},
  \bibinfo{author}{\bibfnamefont{G.}~\bibnamefont{Semerjian}},
  \bibnamefont{and} \bibinfo{author}{\bibfnamefont{F.}~\bibnamefont{Zamponi}},
  \bibinfo{journal}{Physical review letters} \textbf{\bibinfo{volume}{104}},
  \bibinfo{pages}{207206} (\bibinfo{year}{2010}).

\bibitem[{\citenamefont{Bapst et~al.}(2013)\citenamefont{Bapst, Foini,
  Krzakala, Semerjian, and Zamponi}}]{bapst2013quantum}
\bibinfo{author}{\bibfnamefont{V.}~\bibnamefont{Bapst}},
  \bibinfo{author}{\bibfnamefont{L.}~\bibnamefont{Foini}},
  \bibinfo{author}{\bibfnamefont{F.}~\bibnamefont{Krzakala}},
  \bibinfo{author}{\bibfnamefont{G.}~\bibnamefont{Semerjian}},
  \bibnamefont{and} \bibinfo{author}{\bibfnamefont{F.}~\bibnamefont{Zamponi}},
  \bibinfo{journal}{Physics Reports} \textbf{\bibinfo{volume}{523}},
  \bibinfo{pages}{127} (\bibinfo{year}{2013}).

\bibitem[{\citenamefont{Folena et~al.}(2020)\citenamefont{Folena, Franz, and
  Ricci-Tersenghi}}]{folena2020rethinking}
\bibinfo{author}{\bibfnamefont{G.}~\bibnamefont{Folena}},
  \bibinfo{author}{\bibfnamefont{S.}~\bibnamefont{Franz}}, \bibnamefont{and}
  \bibinfo{author}{\bibfnamefont{F.}~\bibnamefont{Ricci-Tersenghi}},
  \bibinfo{journal}{Physical Review X} \textbf{\bibinfo{volume}{10}},
  \bibinfo{pages}{031045} (\bibinfo{year}{2020}).

\bibitem[{\citenamefont{Mannelli et~al.}(2020)\citenamefont{Mannelli, Biroli,
  Cammarota, Krzakala, Urbani, and Zdeborov{\'a}}}]{mannelli2020marvels}
\bibinfo{author}{\bibfnamefont{S.~S.} \bibnamefont{Mannelli}},
  \bibinfo{author}{\bibfnamefont{G.}~\bibnamefont{Biroli}},
  \bibinfo{author}{\bibfnamefont{C.}~\bibnamefont{Cammarota}},
  \bibinfo{author}{\bibfnamefont{F.}~\bibnamefont{Krzakala}},
  \bibinfo{author}{\bibfnamefont{P.}~\bibnamefont{Urbani}}, \bibnamefont{and}
  \bibinfo{author}{\bibfnamefont{L.}~\bibnamefont{Zdeborov{\'a}}},
  \bibinfo{journal}{Physical Review X} \textbf{\bibinfo{volume}{10}},
  \bibinfo{pages}{011057} (\bibinfo{year}{2020}).

\bibitem[{\citenamefont{Mannelli and
  Zdeborov{\'a}}(2020)}]{mannelli2020thresholds}
\bibinfo{author}{\bibfnamefont{S.~S.} \bibnamefont{Mannelli}} \bibnamefont{and}
  \bibinfo{author}{\bibfnamefont{L.}~\bibnamefont{Zdeborov{\'a}}},
  \bibinfo{journal}{Journal of Statistical Mechanics: Theory and Experiment}
  \textbf{\bibinfo{volume}{2020}}, \bibinfo{pages}{034004}
  (\bibinfo{year}{2020}).

\bibitem[{\citenamefont{Semerjian and
  Monasson}(2003)}]{semerjian2003relaxation}
\bibinfo{author}{\bibfnamefont{G.}~\bibnamefont{Semerjian}} \bibnamefont{and}
  \bibinfo{author}{\bibfnamefont{R.}~\bibnamefont{Monasson}},
  \bibinfo{journal}{Physical Review E} \textbf{\bibinfo{volume}{67}},
  \bibinfo{pages}{066103} (\bibinfo{year}{2003}).

\bibitem[{\citenamefont{Semerjian and
  Weigt}(2004)}]{semerjian2004approximation}
\bibinfo{author}{\bibfnamefont{G.}~\bibnamefont{Semerjian}} \bibnamefont{and}
  \bibinfo{author}{\bibfnamefont{M.}~\bibnamefont{Weigt}},
  \bibinfo{journal}{Journal of Physics A: Mathematical and General}
  \textbf{\bibinfo{volume}{37}}, \bibinfo{pages}{5525} (\bibinfo{year}{2004}).

\bibitem[{\citenamefont{Selman et~al.}(1993)\citenamefont{Selman, Kautz, Cohen
  et~al.}}]{selman1993local}
\bibinfo{author}{\bibfnamefont{B.}~\bibnamefont{Selman}},
  \bibinfo{author}{\bibfnamefont{H.~A.} \bibnamefont{Kautz}},
  \bibinfo{author}{\bibfnamefont{B.}~\bibnamefont{Cohen}},
  \bibnamefont{et~al.}, \bibinfo{journal}{Cliques, coloring, and
  satisfiability} \textbf{\bibinfo{volume}{26}}, \bibinfo{pages}{521}
  (\bibinfo{year}{1993}).

\bibitem[{\citenamefont{Montanari and Semerjian}(2006)}]{montanari2006dynamics}
\bibinfo{author}{\bibfnamefont{A.}~\bibnamefont{Montanari}} \bibnamefont{and}
  \bibinfo{author}{\bibfnamefont{G.}~\bibnamefont{Semerjian}},
  \bibinfo{journal}{Journal of statistical physics}
  \textbf{\bibinfo{volume}{124}}, \bibinfo{pages}{103} (\bibinfo{year}{2006}).

\bibitem[{\citenamefont{Semerjian}(2008)}]{semerjian2008freezing}
\bibinfo{author}{\bibfnamefont{G.}~\bibnamefont{Semerjian}},
  \bibinfo{journal}{Journal of Statistical Physics}
  \textbf{\bibinfo{volume}{130}}, \bibinfo{pages}{251} (\bibinfo{year}{2008}).

\bibitem[{\citenamefont{Barthel et~al.}(2003)\citenamefont{Barthel, Hartmann,
  and Weigt}}]{barthel2003solving}
\bibinfo{author}{\bibfnamefont{W.}~\bibnamefont{Barthel}},
  \bibinfo{author}{\bibfnamefont{A.~K.} \bibnamefont{Hartmann}},
  \bibnamefont{and} \bibinfo{author}{\bibfnamefont{M.}~\bibnamefont{Weigt}},
  \bibinfo{journal}{Physical Review E} \textbf{\bibinfo{volume}{67}},
  \bibinfo{pages}{066104} (\bibinfo{year}{2003}).

\bibitem[{\citenamefont{Guidetti and Young}(2011)}]{guidetti2011complexity}
\bibinfo{author}{\bibfnamefont{M.}~\bibnamefont{Guidetti}} \bibnamefont{and}
  \bibinfo{author}{\bibfnamefont{A.}~\bibnamefont{Young}},
  \bibinfo{journal}{Physical Review E} \textbf{\bibinfo{volume}{84}},
  \bibinfo{pages}{011102} (\bibinfo{year}{2011}).

\bibitem[{\citenamefont{Wang et~al.}(2015)\citenamefont{Wang, Machta, and
  Katzgraber}}]{wang2015population}
\bibinfo{author}{\bibfnamefont{W.}~\bibnamefont{Wang}},
  \bibinfo{author}{\bibfnamefont{J.}~\bibnamefont{Machta}}, \bibnamefont{and}
  \bibinfo{author}{\bibfnamefont{H.~G.} \bibnamefont{Katzgraber}},
  \bibinfo{journal}{Physical Review E} \textbf{\bibinfo{volume}{92}},
  \bibinfo{pages}{063307} (\bibinfo{year}{2015}).

\bibitem[{\citenamefont{Kadowaki and Nishimori}(1998)}]{kadowaki1998quantum}
\bibinfo{author}{\bibfnamefont{T.}~\bibnamefont{Kadowaki}} \bibnamefont{and}
  \bibinfo{author}{\bibfnamefont{H.}~\bibnamefont{Nishimori}},
  \bibinfo{journal}{Physical Review E} \textbf{\bibinfo{volume}{58}},
  \bibinfo{pages}{5355} (\bibinfo{year}{1998}).

\bibitem[{\citenamefont{Ambainis and Regev}(2004)}]{ambainis2004elementary}
\bibinfo{author}{\bibfnamefont{A.}~\bibnamefont{Ambainis}} \bibnamefont{and}
  \bibinfo{author}{\bibfnamefont{O.}~\bibnamefont{Regev}},
  \bibinfo{journal}{arXiv preprint quant-ph/0411152}  (\bibinfo{year}{2004}).

\bibitem[{\citenamefont{Messiah}(1962)}]{messiah1962quantum}
\bibinfo{author}{\bibfnamefont{A.}~\bibnamefont{Messiah}},
  \emph{\bibinfo{title}{Quantum mechanics: volume II}}
  (\bibinfo{publisher}{North-Holland Publishing Company Amsterdam},
  \bibinfo{year}{1962}).

\bibitem[{\citenamefont{Laumann et~al.}(2012)\citenamefont{Laumann, Moessner,
  Scardicchio, and Sondhi}}]{laumann2012quantum}
\bibinfo{author}{\bibfnamefont{C.}~\bibnamefont{Laumann}},
  \bibinfo{author}{\bibfnamefont{R.}~\bibnamefont{Moessner}},
  \bibinfo{author}{\bibfnamefont{A.}~\bibnamefont{Scardicchio}},
  \bibnamefont{and} \bibinfo{author}{\bibfnamefont{S.~L.}
  \bibnamefont{Sondhi}}, \bibinfo{journal}{Physical review letters}
  \textbf{\bibinfo{volume}{109}}, \bibinfo{pages}{030502}
  (\bibinfo{year}{2012}).

\bibitem[{\citenamefont{Dormand and Prince}(1980)}]{dormand1980family}
\bibinfo{author}{\bibfnamefont{J.~R.} \bibnamefont{Dormand}} \bibnamefont{and}
  \bibinfo{author}{\bibfnamefont{P.~J.} \bibnamefont{Prince}},
  \bibinfo{journal}{Journal of computational and applied mathematics}
  \textbf{\bibinfo{volume}{6}}, \bibinfo{pages}{19} (\bibinfo{year}{1980}).

\bibitem[{\citenamefont{Weinberg and Bukov}(2017)}]{weinberg2017quspin}
\bibinfo{author}{\bibfnamefont{P.}~\bibnamefont{Weinberg}} \bibnamefont{and}
  \bibinfo{author}{\bibfnamefont{M.}~\bibnamefont{Bukov}},
  \bibinfo{journal}{SciPost Phys.} \textbf{\bibinfo{volume}{2}},
  \bibinfo{pages}{003} (\bibinfo{year}{2017}),
  \urlprefix\url{https://scipost.org/10.21468/SciPostPhys.2.1.003}.

\bibitem[{\citenamefont{Kowalsky et~al.}(2021)\citenamefont{Kowalsky, Albash,
  Hen, and Lidar}}]{USCchallenge}
\bibinfo{author}{\bibfnamefont{M.}~\bibnamefont{Kowalsky}},
  \bibinfo{author}{\bibfnamefont{T.}~\bibnamefont{Albash}},
  \bibinfo{author}{\bibfnamefont{I.}~\bibnamefont{Hen}}, \bibnamefont{and}
  \bibinfo{author}{\bibfnamefont{D.}~\bibnamefont{Lidar}}, \bibinfo{journal}{in
  preparation}  (\bibinfo{year}{2021}).

\bibitem[{\citenamefont{Bernaschi et~al.}(2021)\citenamefont{Bernaschi, Bisson,
  Fatica, Marinari, Martin-Mayor, Parisi, and Ricci-Tersenghi}}]{leadingXORSAT}
\bibinfo{author}{\bibfnamefont{M.}~\bibnamefont{Bernaschi}},
  \bibinfo{author}{\bibfnamefont{M.}~\bibnamefont{Bisson}},
  \bibinfo{author}{\bibfnamefont{M.}~\bibnamefont{Fatica}},
  \bibinfo{author}{\bibfnamefont{E.}~\bibnamefont{Marinari}},
  \bibinfo{author}{\bibfnamefont{V.}~\bibnamefont{Martin-Mayor}},
  \bibinfo{author}{\bibfnamefont{G.}~\bibnamefont{Parisi}}, \bibnamefont{and}
  \bibinfo{author}{\bibfnamefont{F.}~\bibnamefont{Ricci-Tersenghi}},
  \bibinfo{journal}{arXiv preprint arXiv:2102.09510}  (\bibinfo{year}{2021}).

\bibitem[{\citenamefont{Young et~al.}(2010)\citenamefont{Young, Knysh, and
  Smelyanskiy}}]{young2010first}
\bibinfo{author}{\bibfnamefont{A.}~\bibnamefont{Young}},
  \bibinfo{author}{\bibfnamefont{S.}~\bibnamefont{Knysh}}, \bibnamefont{and}
  \bibinfo{author}{\bibfnamefont{V.}~\bibnamefont{Smelyanskiy}},
  \bibinfo{journal}{Physical review letters} \textbf{\bibinfo{volume}{104}},
  \bibinfo{pages}{020502} (\bibinfo{year}{2010}).

\end{thebibliography}

\end{document}